\documentclass[
 aps,
 pra,
 reprint,
 amsmath,amssymb,
 superscriptaddress,
 floatfix,
 longbibliography,
]{revtex4-2}

\usepackage{graphicx}
\usepackage{dcolumn}
\usepackage{bm}
\usepackage{dsfont}
\usepackage[caption=false]{subfig}
\usepackage{mathtools,slashed}
\usepackage{mathrsfs}
\usepackage{microtype}

\allowdisplaybreaks
\usepackage[colorlinks=true,linkcolor=blue,citecolor=blue,urlcolor=blue]{hyperref}

\graphicspath{{figures/}{./}}

\newcommand{\bk}[1][]{\mathbf{k}{#1}}

\newcommand{\tsb}[1]{_{\text{#1}}}

\newcommand{\cop}[1][]{\hat{a}^{\dagger}_{\mathbf{k}{#1},\lambda{#1}}}
\newcommand{\aop}[1][]{\hat{a}_{\mathbf{k}{#1},\lambda{#1}}}

\begin{document}

\title{Time-resolved transient absorption spectroscopy with quantum light}

\author{Mingrui Yuan}
\affiliation{Wyant College of Optical Sciences, University of Arizona, Tucson, AZ 85721}
\affiliation{Department of Physics, University of Arizona, Tucson, AZ 85721}

\author{Nikolay V. Golubev}
\email{ngolubev@arizona.edu}
\affiliation{Department of Physics, University of Arizona, Tucson, AZ 85721}

\date{\today}

\begin{abstract}
We present a fully quantum theory of time-resolved transient absorption spectroscopy, providing a general framework for describing the linear response of a nonequilibrium target to quantum probe fields beyond the semiclassical approximation. We elucidate how quantum correlations between probe-field frequency modes couple to coherences in the target and derive a simple expression for computing the resulting spectroscopic signals. We identify conditions under which quantum light can modify transient absorption spectra beyond the limits achievable with classical fields. Our work establishes a theoretical foundation for exploiting quantum light in ultrafast spectroscopy and suggests new routes for resolving dynamics in matter on timescales shorter than the duration of the probe pulse.
\end{abstract}

\maketitle

\section{Introduction}
\label{sec:introduction}

Time-resolved transient absorption spectroscopy (TR-TAS) has become one of the central tools for observing the evolution of matter in real time~\cite{goulielmakis2010,chini2013,wei2017,kobayashi2017,sabbar2017,geneaux2019,timmers2019,driver2020,kobayashi2020b,rott2021,matselyukh2022,wang2024,matselyukh2025,chakraborty2026,bai2026,volkov2026}. In this technique, a sample in a nonequilibrium state is illuminated by a probe laser pulse whose arrival time is varied relative to the ongoing sample's dynamics. By measuring the transmitted probe spectrum as a function of the time delay, one can track the evolution of the sample's optical response and gain insight into its underlying quantum dynamics.

Most existing theoretical formulations of TR-TAS~\cite{gaarde2011,santra2011,pabst2012,perfetto2015,baekhoj2016,wu2016,kolbasova2019,chen2020TA,golubev2021,yuen2024} operate in a semiclassical regime, i.e. they treat matter as a quantum object while describing light classically. This semiclassical description of light-matter interactions was well justified for conventional coherent fields used in linear absorption spectroscopy, since the essential field observables closely approximate those of classical electromagnetic waves, particularly in the limit of large photon numbers. While some aspects of TR-TAS have also been investigated starting from a general quantum formulation of light-matter interactions, including covariance-based transient absorption signals with stochastic probe fields~\cite{osipov2019} and connections between absorption and Raman-like processes~\cite{cavaletto2023}, a general quantum formulation of TR-TAS remains lacking.

A fully quantum description of light and its interaction with matter become increasingly relevant as advances in quantum optics enable the generation and control of light with tailored quantum properties~\cite{breitenbach1997,ourjoumtsev2006,bimbard2010,cooper2013ql,chekhova2015,gorlach2020,gorlach2023,rasputnyi2024,sennary2025light,kern2026}. Unlike classical-like coherent pulses, quantum states of light can exhibit nonclassical photon statistics and correlations, introducing additional degrees of freedom for controlling light-matter interactions. These properties have attracted considerable interest in nonlinear spectroscopy~\cite{dorfman2014,dorfman2016,mukamel2020,dorfman2026,fan2026}, where the quantum nature of light has been considered since the early development of the field~\cite{mukamel1995book}. This is a consequence of the fact that nonlinear optical signals are particularly sensitive to quantum properties of light because they depend naturally on higher-order field correlations. One prominent example is two-photon absorption spectroscopy, where quantum correlations between photons can modify excitation probabilities and spectral selectivity, leading to responses that differ from those obtained with classical light~\cite{dayan2007,schlawin2017,schlawin2018}. 

Although the possibilities for manipulating linear spectroscopic signals through quantum properties of light are more limited than in nonlinear spectroscopy, they are not absent. In particular, the correlations between distinct frequency modes of a probe pulse can influence how the evolving quantum state of the target is encoded in the transient absorption spectrum. As we demonstrate in this work, these correlations provide a means of controlling the spectral signatures of nonequilibrium dynamics even within the linear-response regime.

Here, we develop a fully quantum theory of linear TR-TAS in which light and matter are treated on the same footing. Within this framework, the transient absorption signal is expressed explicitly in terms of the quantum correlations between frequency modes of the applied probe field and the corresponding correlations between energy levels of the target. By establishing a direct connection between the properties of the probe field and the target, we identify characteristic energy-level configurations of the target and determine how the probe photons couple these levels. We show that, for specific configurations, quantum correlations between field modes can be engineered to surpass the bounds achievable with classical light, providing new opportunities to control transient absorption signals. We illustrate our findings with several realistic numerical examples, demonstrating how incoherent, coherent, and squeezed-light probes influence time-resolved transient absorption spectra.

The article is organized as follows. Section~\ref{sec:theory} introduces the theoretical framework, describing the system, the quantized field, and the detected observable. In Sec.~\ref{sec:linear_nonlinear}, we identify the three contributions to the detected photon number and discuss the class of spectroscopic measurements associated with each of them. Section~\ref{sec:TAS} specializes the formalism to TR-TAS and derives the central expressions of this work. Section~\ref{sec:applications} contains the results of our numerical simulations, comparing incoherent and coherent probes in Sec.~\ref{sec:probe_spectral_coherence} and coherent and squeezed probes in Sec.~\ref{sec:sqp-picture}. We summarize our results and conclude the paper in Sec.~\ref{sec:conclusions}. Technical details of the derivations are given in the Appendices.

\section{Theoretical framework}
\label{sec:theory}

In a typical optical spectroscopy experiment, a sample is illuminated with electromagnetic radiation, and the transmitted, reflected, or emitted light is measured by a photodetector or a similar device. The primary experimental observable is the intensity of the optical field emerging from the sample, which is typically dispersed using a diffraction grating and recorded as a function of the photon energy. Assuming ideal dispersion, such that each detector pixel corresponds to a well-defined photon energy $\omega$, and using the definition of intensity as the optical power $P$ distributed over the detector area $A$, the measured spectral intensity can be expressed as
\begin{equation}
\label{eq:Iw_exp}
    I(\omega)
    = \frac{P}{A}
    = \frac{N_{\omega}\hbar\omega}{A t_{\text{e}}},
\end{equation}
where $N_{\omega}$ denotes the number of photons detected at energy $\omega$ during the exposure time $t_{\text{e}}$.

The expression Eq.~(\ref{eq:Iw_exp}) establishes a direct connection between experimentally measured intensity and the quantum description of the electromagnetic field. In particular, the detected photon number $N_{\omega}$ corresponds to the occupation of the radiation modes coupled to the detector and can be expressed in terms of expectation values of photon number operators within a QED framework, as we show in the next sections. In the following, we use atomic units for convenience, with $\hbar=e=m_{\mathrm{e}}=4\pi\epsilon_{0}=1$ and $c=1/\alpha_{\text{fs}}$, where $\alpha_{\text{fs}}$ is the fine-structure constant.

\subsection{Description of the system}
The total Hamiltonian of the system, describing the target, the electromagnetic field, and their interaction, is written in the Schr\"odinger picture as
\begin{equation}
\label{eq:H_total}
    \hat{H}=\hat{H}_0+\hat{\mathcal{V}},
\end{equation}
where
\begin{equation}
\label{eq:H_0}
    \hat{H}_0=\hat{H}_{\text{t}}+\hat{H}_{\text{b}},
\end{equation}
is the operator composed of the non-interacting target, $\hat{H}_{\text{t}}$, and the probe beam, $\hat{H}_{\text{b}}$, Hamiltonians, while $\hat{\mathcal{V}}$ term describes their coupling. In the most general formulation, these Hamiltonians can be explicitly time dependent, allowing for externally driven targets and probe beams that undergo externally induced modifications during propagation. Unless stated otherwise, we restrict our discussion to time-independent Hamiltonians, where the dynamics originate from coherent superpositions of the target and probe-beam states. The interaction Hamiltonian then governs the exchange of populations and coherences between these states.

The quantum state of the total system is described by the density operator $\hat{\rho}(t)$, whose evolution is governed by the von Neumann equation
\begin{equation}
\label{eq:von_neumann}
    i\frac{\partial\hat{\rho}(t)}{\partial t} =
    \left[\hat{H},\hat{\rho}(t)\right].
\end{equation}
The formal solution of Eq.~(\ref{eq:von_neumann}) is
\begin{equation}
\label{eq:rho_solution}
    \hat{\rho}(t) =
    \hat{U}(t,t_i)\hat{\rho}(t_i)\hat{U}^{\dagger}(t,t_i),
\end{equation}
where $\hat{U}(t,t_i)$ is the time-evolution operator generated by the total Hamiltonian, $t_i$ is the initial time, and $\hat{\rho}(t_i)$ is the corresponding initial density operator. Assuming that the target and the probe beam are uncorrelated at $t_i$, the total density operator can be written as
\begin{equation}
\label{eq:rho_ti}
    \hat{\rho}(t_i) =
    \hat{\rho}_{\text{t}}(t_i)\otimes\hat{\rho}_{\text{b}}(t_i),
\end{equation}
where $\hat{\rho}_{\text{t}}$ and $\hat{\rho}_{\text{b}}$ describe the initial quantum states of the target and probe beam, respectively. Although the initial density operator factorizes into independent target and beam contributions, their interaction generally leads to the development of correlations, and the density operator at later times can no longer be expressed as a simple tensor product of the individual subsystem density operators.

\subsection{Quantum light model and light-matter interaction}
Within the QED framework, the probe beam is described as a set of quantized field modes, each characterized by its wave vector $\bk$, polarization index $\lambda$, and the frequency $\omega_{\bk} = |\bk|/\alpha\tsb{fs}$. The corresponding free-field Hamiltonian is therefore given by
\begin{equation}
\label{eq:Hb}
	\hat{H}\tsb{b} = \sum_{\bk,\lambda} \omega_{\bk} 
	\left(
		\cop \aop + \frac{1}{2}
	\right),
\end{equation}
where $\cop{}$ and $\aop{}$ are the creation and annihilation operators of the field mode $(\bk,\lambda)$, respectively, satisfying the standard bosonic commutation relations. 

The interaction between the target and the quantized radiation field is described by the minimal coupling Hamiltonian
\begin{equation}
\label{eq:H_int}
    \hat{\mathcal{V}} =
    \int d^3r\,
    \left[
        -\alpha_{\text{fs}}\,
        \hat{\mathbf{j}}(\mathbf{r}) \cdot \hat{\mathbf{A}}(\mathbf{r})
        + \frac{\alpha_{\mathrm{fs}}^2}{2}\, \hat{n}(\mathbf{r}) \hat{\mathbf{A}}^2(\mathbf{r})
    \right].
\end{equation}
The first and second terms describe the paramagnetic and diamagnetic couplings, respectively. Here, $\hat{\mathbf{A}}(\mathbf{r})$ is the quantized vector potential operator defined as
\begin{equation}
\label{eq:A_op}
    \hat{\mathbf{A}}(\mathbf{r})=
    \sum_{\mathbf{k}, \lambda} \sqrt{\frac{2 \pi}{V \omega_{\mathbf{k}} \alpha_{\text{fs}}^2}}
    \left(
        \hat{a}_{\mathbf{k}, \lambda} 
        \boldsymbol{\epsilon}_{\mathbf{k}, \lambda} 
        e^{i \mathbf{k} \cdot \mathbf{r}}
        +
        \hat{a}_{\mathbf{k}, \lambda}^{\dagger} 
        \boldsymbol{\epsilon}_{\mathbf{k}, \lambda}^* 
        e^{-i \mathbf{k} \cdot \mathbf{r}}
    \right),
\end{equation}
where $\boldsymbol{\epsilon}_{\mathbf{k}, \lambda}$ is the polarization vector associated with the field mode $(\mathbf{k},\lambda)$. The operator $\hat{\mathbf{j}}(\mathbf{r})$ denotes the field-free current density operator,
\begin{equation}
\label{eq:j_operator}
    \hat{\mathbf{j}}(\mathbf{r})=
    -\frac{i}{2} \left[
        \hat{\psi}^{\dagger}(\mathbf{r})
        \boldsymbol{\nabla}\hat{\psi}(\mathbf{r})
        - \left( \boldsymbol{\nabla}\hat{\psi}^{\dagger}(\mathbf{r}) \right) \hat{\psi}(\mathbf{r})
    \right],
\end{equation}
while $\hat{n}(\mathbf{r})$ is the target density operator,
\begin{equation}
\label{eq:n_operator}
    \hat{n}(\mathbf{r}) =
    \hat{\psi}^{\dagger}(\mathbf{r})
    \hat{\psi}(\mathbf{r}).
\end{equation}
Here, $\hat{\psi}^{\dagger}(\mathbf{r})$ and $\hat{\psi}(\mathbf{r})$ are the field creation and annihilation operators, respectively, of the target, which satisfy the standard fermionic anticommutation relations.

\subsection{Description of the observable}
To describe the measurement, we introduce a quantum-mechanical observable $\hat{O}$ and evaluate its expectation value with respect to the density matrix of the system:
\begin{equation}
\label{eq:Oexpectation}
    \langle \hat{O}(t_f)\rangle = 
    \operatorname{Tr}\left[ \hat{O} \hat{\rho}(t_f) \right],
\end{equation}
where $t_f$ denotes the time at which the measurement is performed. For the photon-counting measurement, the relevant observable is an effective photon-number operator, which can be written as
\begin{equation}
    \hat{O} =
    \sum_\lambda \int d^3 k ~
    W(\mathbf{k})~
    \hat{a}_{\mathbf{k}, \lambda}^\dagger
    \hat{a}_{\mathbf{k}, \lambda},
\end{equation}
where $W(\mathbf{k})$ is a detector response function that specifies the field modes contributing to the measurement. In particular, $W(\mathbf{k})$ can account for the geometrical location and acceptance angle of the detector, as well as its spectral and polarization response. Experimentally, varying the detector position provides a means of distinguishing photons propagating along the incident direction from those emitted or scattered into other directions, thereby enabling the separation of transmitted and scattered radiation.

For example, consider a detector that is insensitive to the polarization of the field, has perfect frequency resolution, and is positioned along the direction of $\mathbf{k}'$, thereby selecting photons characterized by the wave vector $\mathbf{k}'_{\omega}=\alpha_{\text{fs}} \omega\ \mathbf{k}' / |\mathbf{k}'|$. In the idealized limit of infinitesimal angular acceptance, the corresponding detector response function can be written as
\begin{equation}
    W(\mathbf{k}) =
    \delta^{(3)}
    \left( \mathbf{k} - \mathbf{k}'_{\omega} \right),
\end{equation}
and the photon-number observable then becomes
\begin{equation}
\label{eq:O_kw}
    \hat{O}_{\mathbf{k}'_{\omega}} = \sum_\lambda
        \hat{a}_{\mathbf{k}'_{\omega}, \lambda}^\dagger
        \hat{a}_{\mathbf{k}'_{\omega}, \lambda}.
\end{equation}

Substituting the photon-number observable from Eq.~(\ref{eq:O_kw}) into Eq.~(\ref{eq:Oexpectation}), the number of photons $N_{\omega}$ detected at frequency $\omega$ can be expressed in terms of the density matrix at the measurement time $t_f$. Using the unitary time evolution of the density matrix, Eq.~(\ref{eq:rho_solution}), and the cyclic property of the trace, this expression can equivalently be written in the Heisenberg picture as
\begin{widetext}
\begin{equation}
\label{eq:Nw}
\begin{split}
    N_{\omega} & = \sum_\lambda \operatorname{Tr}
        \left[ 
            \hat{a}_{\mathbf{k}'_{\omega}, \lambda}^\dagger
            \hat{a}_{\mathbf{k}'_{\omega}, \lambda} 
            \hat{\rho}(t_f) 
        \right]
    = \sum_\lambda \operatorname{Tr}
        \left[ 
            \hat{a}_{\mathbf{k}'_{\omega}, \lambda}^\dagger
            \hat{a}_{\mathbf{k}'_{\omega}, \lambda} 
            \hat{U}(t_f,t_i)\hat{\rho}(t_i)\hat{U}^{\dagger}(t_f,t_i)
        \right] \\
    & = \sum_\lambda \operatorname{Tr}
        \left[
            \hat{U}^{\dagger}(t_f,t_i)
            \hat{a}_{\mathbf{k}'_{\omega}, \lambda}^\dagger
            \hat{U}(t_f,t_i)
            \hat{U}^{\dagger}(t_f,t_i)
            \hat{a}_{\mathbf{k}'_{\omega}, \lambda} 
            \hat{U}(t_f,t_i)
            \hat{\rho}(t_i)
        \right]
    = \sum_\lambda \operatorname{Tr}
        \left[
            \hat{a}_{\mathbf{k}'_{\omega}, \lambda}^\dagger(t_f)
            \hat{a}_{\mathbf{k}'_{\omega}, \lambda}(t_f)
            \hat{\rho}(t_i)
        \right].
\end{split}
\end{equation}
\end{widetext}

The time evolution of the annihilation and creation operators, $\hat{a}_{\mathbf{k}'_{\omega}, \lambda}(t)$ and $\hat{a}_{\mathbf{k}'_{\omega}, \lambda}^\dagger(t)$, is governed by the Heisenberg equation of motion and is determined by the full Hamiltonian of the coupled target-field system. In Appendix~\ref{app:field_opp}, we derive explicit expressions for these operators in terms of the target current. In particular, the annihilation operator can be written as
\begin{equation}
\label{eq:a_decomp}
    \hat{a}_{\mathbf{k}'_{\omega},\lambda}(t) = 
        e^{-i\omega(t-t_i)} \hat{a}_{\mathbf{k}'_{\omega},\lambda}
        + \hat{a}^{\text{r}}_{\mathbf{k}'_{\omega},\lambda}(t),
\end{equation}
where the first term describes the free evolution of the initially populated field mode, while $\hat{a}^{\text{r}}_{\mathbf{k}'_{\omega},\lambda}(t)$ describes the radiation generated by the target:
\begin{equation}
\label{eq:a_rad}
\begin{split}
    \hat{a}^{\text{r}}_{\mathbf{k}'_{\omega},\lambda}(t) = 
        i\sqrt{\frac{2\pi}{V\omega}}
        & \int_{t_i}^{t} dt'\,
        e^{-i\omega(t-t')} \\
        \times & \int d^3r\,
        \boldsymbol{\epsilon}_{\mathbf{k}'_{\omega},\lambda}^{*}
        \cdot
        \hat{\mathbf{J}}(\mathbf{r},t')
        e^{-i\mathbf{k}'_{\omega}\cdot\mathbf{r}}.
\end{split}
\end{equation}
Here, $\hat{\mathbf{J}}(\mathbf{r},t)$ denotes the Heisenberg-picture form of the total current density operator of the target under the influence of the field defined as
\begin{equation}
    \hat{\mathbf{J}}(\mathbf{r},t) =
    \hat{U}^{\dagger}(t,t_i)
    \left[ \,
        \hat{\mathbf{j}}(\mathbf{r}) -
        \alpha_{\mathrm{fs}}
        \hat{\mathbf{A}}(\mathbf{r})
        \hat{n}(\mathbf{r})
    \right]
    \hat{U}(t,t_i).
\end{equation}
Thus, the radiation emitted by the target is fully determined by the time-dependent total current density of the coupled target-field system, providing the connection between the microscopic target dynamics and the quantized radiation field.

\section{Linear and nonlinear optical spectroscopy}
\label{sec:linear_nonlinear}

We now use the explicit form of the field operators derived above to examine the structure of the detected photon number $N_{\omega}$ and to identify the distinct physical contributions underlying linear and nonlinear optical spectroscopies. Substituting the decomposition of the annihilation operator, Eq.~(\ref{eq:a_decomp}), together with its Hermitian conjugate, Eq.~(\ref{eq:a_decomp_conj}), derived in Appendix~\ref{app:field_opp}, into the expression for $N_{\omega}$, Eq.~(\ref{eq:Nw}), we obtain
\begin{equation}
    N_\omega = N_{\omega}^{\text{bg}} + N_{\omega}^{\text{int}} + N_{\omega}^{\text{rad}}.
\end{equation}
This decomposition follows directly from the separation of the field operator into its freely evolving and target-generated parts. It provides a useful distinction between the contribution of the incident field, the radiation generated by the target, and their mutual interference. The relative importance of these contributions depends not only on the incident radiation and the target response, but also on the detector position, which determines the wave vector $\mathbf{k}'_{\omega}$ of the detected modes. In particular, the detection geometry can suppress or exclude certain contributions from the measured signal. By varying the detector position, one can therefore selectively probe radiation propagating along the incident direction or radiation emitted and scattered into other directions. We consider these contributions and their dependence on the detection geometry separately below.

The background contribution,
\begin{equation}
    N_{\omega}^{\text{bg}} =
    \sum_\lambda \operatorname{Tr}
        \left[ 
            \hat{a}_{\mathbf{k}'_{\omega},\lambda}^{\dagger}
            \hat{a}_{\mathbf{k}'_{\omega},\lambda} \hat{\rho}_{\mathrm{b}}(t_i) 
        \right],
\end{equation}
is particularly simple because the initial density matrix is factorized, Eq.~(\ref{eq:rho_ti}), while the operators $\hat{a}_{\mathbf{k}'_{\omega},\lambda}$ and $\hat{a}_{\mathbf{k}'_{\omega},\lambda}^{\dagger}$ act exclusively on the Hilbert space of the field, so that the trace over the target degrees of freedom reduces to $\operatorname{Tr}[\hat{\rho}_{\text{t}}(t_i)] \equiv 1$. Therefore, $N_{\omega}^{\text{bg}}$ contribution is determined solely by the field initially present in the detected mode and is thus independent of the target dynamics. To record this background signal, the detector must be placed along the incident propagation direction.

The interference contribution,
\begin{equation}
\label{eq:Nw_int}
    N_{\omega}^{\mathrm{int}} =
    2\,\mathrm{Re}\sum_\lambda e^{i\omega(t_f-t_i)} \operatorname{Tr} 
        \left[ 
            \hat{a}_{\mathbf{k}'_{\omega},\lambda}^{\dagger}
            \hat{a}^{\mathrm{r}}_{\mathbf{k}'_{\omega},\lambda}(t_f) 
            \hat{\rho}(t_i) 
        \right],
\end{equation}
arises from the interference between the freely propagating incident field and the field radiated by the target. Because this contribution contains the operator $\hat{a}_{\mathbf{k}'_{\omega},\lambda}^{\dagger}$ of the field at initial moment of time, the detected mode must have a nonzero field amplitude in the same mode at the detection time. Similarly to the background term, this contribution vanishes when the detector is placed away from the incident propagation direction in the idealized limit of a perfectly collimated incident field. Furthermore, the interference term is nonzero only if the detected frequency $\omega$ is present in the incident field, since otherwise there is no incident field component at the detected frequency with which the target-generated radiation can interfere. This frequency-selective nature makes the interference contribution particularly relevant to transmission-based measurements, including absorption and stimulated-emission spectroscopy, where the detected signal arises from the interference between the incident field and the field generated by the target.

The radiation contribution, 
\begin{equation}
\begin{split}
    N_{\omega}^{\mathrm{rad}} =
    \sum_\lambda \operatorname{Tr} 
        \left[ 
            \left\{ \hat{a}^{\mathrm{r}}_{\mathbf{k}'_{\omega},\lambda}(t_f) \right\}^\dagger
            \hat{a}^{\mathrm{r}}_{\mathbf{k}'_{\omega},\lambda}(t_f) 
            \hat{\rho}(t_i) 
        \right],
\end{split}
\end{equation}
describes the photon signal generated by the target itself, either through initial coherences present in the target or through its linear or nonlinear optical response to the probe field. Unlike the background and interference contributions, this term can remain nonzero when the detector is placed away from the incident propagation direction or when the detected frequency is absent from the incident radiation. It therefore provides the basis for background-free detection of target-generated radiation and encompasses a broad range of spectroscopic signals, including spontaneous and stimulated emission, Raman and Rayleigh scattering, four-wave mixing, high-harmonic generation, and other nonlinear optical processes in matter driven by intense laser fields.

\section{Transient absorption spectroscopy}
\label{sec:TAS}

As a first application of the general framework introduced in Secs.~\ref{sec:theory} and~\ref{sec:linear_nonlinear}, we consider TR-TAS. We assume that the target has been prepared in a general nonequilibrium state and take this state as the initial condition for the subsequent dynamics. The preparation process itself is not considered explicitly. Instead, we focus on the response of the prepared target to a weak probe field and on how this response is encoded in the transmitted probe signal.

Within the dipole approximation, the target radiation induced by the probe field is, to linear order in the field, given by
\begin{equation}
\label{eq:ar_t_dip}
\begin{split}
    \hat{a}^{\text{r}}_{\mathbf{k}'_{\omega},\lambda}(t) = 
        & i\sqrt{\frac{2\pi\omega}{V}}
        \int_{t_i}^{t}dt'\,
        \int_{t_i}^{t}dt''\,
        e^{-i\omega(t-t')} \\
        & \times
        \Theta(t'-t'')\,
        \boldsymbol{\epsilon}_{\mathbf{k}'_{\omega},\lambda}^{*}
        \cdot
        \hat{\boldsymbol{\mathcal D}}(t',t'')
        \cdot
        \hat{\mathbf E}_{\mathrm I}(t'').
\end{split}
\end{equation}
Here, $\Theta(t'-t'')$ is the Heaviside step function, $\hat{\mathbf E}_{\mathrm I}(t)$ is the interaction-picture electric-field operator, and $\hat{\boldsymbol{\mathcal D}}(t',t'')$ denotes the dipole-dipole commutator tensor with elements
\begin{equation}
    \hat{\mathcal D}_{ij}(t',t'')
    =
    \left[
        \hat{d}^{i}_{\mathrm I}(t'),
        \hat{d}^{j}_{\mathrm I}(t'')
    \right],
\end{equation}
where $\hat{\mathbf d}_{\mathrm I}(t)= \hat{U}_\text{t}^\dagger(t,t_i)\, \hat{\mathbf{d}}\, \hat{U}_\text{t}(t,t_i)$ is the interaction-picture dipole operator evolved under the unperturbed target Hamiltonian $\hat{H}_{\text{t}}$. A detailed derivation of Eq.~(\ref{eq:ar_t_dip}) is given in Appendix~\ref{app:dip_field_opp}.

Substituting Eq.~(\ref{eq:ar_t_dip}) into the interference contribution to the photon number, Eq.~(\ref{eq:Nw_int}), and rearranging the terms, we obtain
\begin{widetext}
\begin{equation}
\label{eq:Nw_int_TA}
\begin{split}
    N_{\omega}^{\text{int}}(\tau) & = 
    - 2\,\operatorname{Re}
    \bigg[
        i \sum_{\lambda}
        \int_{t_i}^{t_f} dt' \int_{t_i}^{t_f} dt''
        \boldsymbol{\epsilon}_{\mathbf{k}'_{\omega},\lambda}^{*}
        \cdot
        \boldsymbol{\chi}(t',t'')
        \cdot
        \left\langle
            \hat{E}^{(-)}_{\mathbf{k}'_{\omega},\lambda}(t'-\tau)
            \hat{\mathbf E}_{\mathrm I}(t''-\tau)
        \right\rangle
    \bigg],
\end{split}
\end{equation}
\end{widetext}
Here, $\boldsymbol{\chi}(t',t'')$ is the two-time electric susceptibility tensor of the target, which characterizes its linear response to the quantum probe field. Its components are defined by
\begin{equation}
    \chi_{ij}(t',t'')
    =
    -i\Theta(t'-t'')
    \operatorname{Tr}
    \left\{
        \hat{\rho}_{\mathrm t}(t_i)
        \hat{\mathcal D}_{ij}(t',t'')
    \right\},
\end{equation}
where the trace is taken over the target degrees of freedom. In the energy eigenbasis, $\hat H_{\mathrm t}|n\rangle = E_n|n\rangle$, and for a general initial state given by the density operator
\begin{equation}
    \hat{\rho}_{\mathrm t}(t_i) =
    \sum_{m,n} \rho_{nm}
    |n\rangle\langle m|,
\end{equation}
the susceptibility can be written explicitly as
\begin{widetext}
\begin{equation}
\label{eq:suscept}
    \chi_{ij}(t',t'')
    =
    2\Theta(t'-t'')
    \operatorname{Im}
    \left[
        \sum_{m,n,l}
        \rho_{nm}\,
        d^i_{ml}d^j_{ln}
        e^{-i(E_m-E_n)t_i}
        e^{i(E_m-E_l+i\Gamma_l/2)t'}
        e^{i(E_l-E_n-i\Gamma_l/2)t''}
    \right],
\end{equation}
\end{widetext}
where $\rho_{nm}$ are, in general, complex matrix elements specifying the populations and coherences of the initial state, $d^{i}_{ml}$ denotes the $i$-th component of the dipole matrix element, $\mathbf d_{ml}=\langle m|\hat{\mathbf d}|l\rangle$,  and $\Gamma_l$ denotes the decay rate, or equivalently the energy width, associated with the finite lifetime of the state $|l\rangle$. A detailed derivation of this expression is given in Appendix~\ref{app:susceptibility}.

The second non-trivial factor in Eq.~(\ref{eq:Nw_int_TA}) is the mode-resolved field correlation function,
\begin{equation}
\begin{split}
    \Big\langle
        \hat{E}^{(-)}_{\mathbf{k}'_{\omega},\lambda} & (t'-\tau)
        \hat{\mathbf E}_{\mathrm I}(t''-\tau)
    \Big\rangle = \\
    & \operatorname{Tr}
    \left\{
        \hat{\rho}_{b}(t_i)
        \hat{E}^{(-)}_{\mathbf{k}'_{\omega},\lambda}(t'-\tau)
        \hat{\mathbf E}_{\mathrm I}(t''-\tau)
    \right\},
\end{split}
\end{equation}
where we introduced a delay parameter $\tau$ that defines the temporal position of the probe field relative to the target evolution. The negative-frequency component of the electric-field operator,
\begin{equation}
\label{eq:E_I_minus}
    \hat E^{(-)}_{\mathbf{k},\lambda}(t)
    =
    -i \sqrt{\frac{2\pi\omega_{\mathbf{k}}}{V}}\,
    \hat{\mathrm a}_{\mathbf{k},\lambda}^{\dagger}
    e^{i\omega_{\mathbf{k}}(t-t_i)},
\end{equation}
together with its positive frequency component, $\hat{E}^{(+)}_{\mathbf{k},\lambda}(t) = \left[\hat{E}^{(-)}_{\mathbf{k},\lambda}(t)\right]^{\dagger}$, constitute the standard expansion of the interaction-picture electric-field operator,
\begin{equation}
\label{eq:E_I_decomp}
    \hat{\mathbf E}_{\mathrm I}(t)
    =
    \sum_{\mathbf{k},\lambda}
    \left[
        \boldsymbol{\epsilon}_{\mathbf{k},\lambda}
        \hat E^{(+)}_{\mathbf{k},\lambda}(t)
        +
        \boldsymbol{\epsilon}_{\mathbf{k},\lambda}^{*}
        \hat E^{(-)}_{\mathbf{k},\lambda}(t)
    \right].
\end{equation}
Substituting these expressions into the field correlation function yields
\begin{widetext}
\begin{equation}
\label{eq:field_corr}
\begin{split}
    \big\langle
        \hat{E}^{(-)}_{\mathbf{k}'_{\omega},\lambda}(t'-\tau)
        & \hat{\mathbf E}_{\mathrm I}(t''-\tau)
    \big\rangle =
    \frac{2\pi}{V}
    \sum_{\mathbf{k},\nu}
    \sqrt{\omega\,\omega_{\mathbf{k}}} \\
    & \Bigg[
        \boldsymbol{\epsilon}_{\mathbf{k},\nu}
        \left\langle
            \hat a_{\mathbf{k}'_{\omega},\lambda}^{\dagger}
            \hat a_{\mathbf{k},\nu}
        \right\rangle
        e^{i\omega t'}e^{-i\omega_{\mathbf{k}}t''} e^{-i(\omega-\omega_{\mathbf{k}})(\tau+t_i)}
      - \boldsymbol{\epsilon}_{\mathbf{k},\nu}^{*}
        \left\langle
            \hat a_{\mathbf{k}'_{\omega},\lambda}^{\dagger}
            \hat a_{\mathbf{k},\nu}^{\dagger}
        \right\rangle
        e^{i\omega t'}e^{i\omega_{\mathbf{k}}t''} e^{-i(\omega+\omega_{\mathbf{k}})(\tau+t_i)}
    \Bigg],
\end{split}
\end{equation}
\end{widetext}
where $\left\langle \hat a_{\mathbf{k}'_{\omega},\lambda}^{\dagger} \hat a_{\mathbf{k},\nu} \right\rangle$ and $\left\langle \hat a_{\mathbf{k}'_{\omega},\lambda}^{\dagger} \hat a_{\mathbf{k},\nu}^{\dagger} \right\rangle$ denote the normal and anomalous field correlations, respectively, between the detected mode $(\mathbf{k}'_{\omega},\lambda)$ and the incident modes $(\mathbf{k},\nu)$. These correlations contain the information about the quantum state of the incident radiation and, in general, depend on the mode indices, frequencies, polarizations, and relative phases of the field modes.

Substituting the spectral representation of the susceptibility, Eq.~(\ref{eq:suscept}), and the field correlation function, Eq.~(\ref{eq:field_corr}), into Eq.~(\ref{eq:Nw_int_TA}), and carrying out the integrations over $t'$ and $t''$ (see Appendix~\ref{app:susceptibility} for details), gives
\begin{widetext}
\begin{equation}
\label{eq:TA_full}
\begin{split}
    N_{\omega}^{\text{int}}(\tau)
    = -\frac{8\pi^2}{V}\sqrt{\omega}\,
    \operatorname{Im}\Bigg\{
    \sum_{i,j}\sum_{m,n}
    \rho_{nm}\,e^{i(E_m-E_n)\tau}
    \sum_l
    \left[
    \frac{d^i_{ml}d^j_{ln}}
         {E_l-E_m-\omega-i\Gamma_l/2}
    +
    \frac{d^j_{ml}d^i_{ln}}
         {E_l-E_n+\omega+i\Gamma_l/2}
    \right]
    \mathcal{P}_{mn}^{ij}(\omega)
    \Bigg\},
\end{split}
\end{equation}
where we introduced the field-dependent spectral factor
\begin{equation}
\label{eq:SF_full}
\begin{split}
    \mathcal{P}_{mn}^{ij}(\omega) =
    \sum_{\lambda,\nu} \sum_\mathbf{k} \sqrt{\omega_{\mathbf{k}}}
        \bigg[
            & \left( \epsilon_{\mathbf{k}'_{\omega},\lambda}^{i} \right)^* \epsilon_{\mathbf{k},\nu}^{j}
            \left\langle \hat a_{\mathbf{k}'_{\omega},\lambda}^{\dagger} \hat a_{\mathbf{k},\nu} \right\rangle
            \delta(\omega-\omega_{\mathbf{k}}+E_m-E_n) \\
            -
            & \left( \epsilon_{\mathbf{k}'_{\omega},\lambda}^{i} \right)^* \left( \epsilon_{\mathbf{k},\nu}^{j} \right)^*
            \left\langle \hat a_{\mathbf{k}'_{\omega},\lambda}^{\dagger} \hat a_{\mathbf{k},\nu}^{\dagger} \right\rangle
            \delta(\omega+\omega_{\mathbf{k}}+E_m-E_n)
        \bigg],
\end{split}
\end{equation}
\end{widetext}
which contains information about the polarization and the correlations between the field modes, and also imposes energy conservation conditions. To derive Eqs.~(\ref{eq:TA_full}) and~(\ref{eq:SF_full}), we utilized the fact that the delta functions constrain the incident field frequency $\omega_{\mathbf{k}}$ to values determined by the detected frequency $\omega$ and the energy difference $E_m-E_n$. In particular, the first delta function imposes $\omega_{\mathbf{k}}=E_m-E_n+\omega$, while the second imposes $\omega_{\mathbf{k}}=E_n-E_m-\omega$. Thus, upon carrying out the corresponding sum over $\mathbf{k}$, the terms that explicitly depend on $\omega_{\mathbf{k}}$ are evaluated at these selected frequencies and can be taken outside the remaining sum. This makes explicit that the detected signal at frequency $\omega$ couples to specific spectral components of the incident field, with the relevant frequency shift determined by the energy difference between the states of the target.

The structure of Eq.~(\ref{eq:TA_full}) closely resembles that of the corresponding semiclassical expression (see, e.g., Ref.~\cite{santra2011}) obtained for a classical incident field. In both cases, the signal is determined by the transition dipole matrix elements connecting the initial and final states with each other, the density matrix elements $\rho_{nm}$ describing the populations and coherences of the initial state, and Lorentzian line shapes whose widths are determined by the lifetimes of the final states. The essential difference is that, in the quantum-field formulation, each spectral line is weighted by the field-dependent spectral factor $\mathcal{P}_{mn}^{ij}(\omega)$. This quantity replaces the corresponding classical electric field spectral profile arising in the corresponding semiclassical expression and encodes the quantum properties of the incident radiation through its mode correlations, polarization structure, and frequency distribution. In particular, for nonclassical states of light such as squeezed or entangled light, these correlations can modify the relative weighting of individual spectral lines in a manner that has no direct classical-field analogue.

\begin{figure}[t]
	\includegraphics[width=8.5cm]{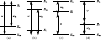}
\caption{Diagrams illustrating possible arrangements of the energy levels and the corresponding energy conservation conditions imposed by Eqs.~(\ref{eq:TA_full}) and~(\ref{eq:SF_full}). The initially populated states $|m\rangle$ and $|n\rangle$ are connected by the coherence $\rho_{nm}$, $|l\rangle$ is the state reached through the two dipole transitions, $\omega$ is the detected photon energy, and $\omega_{\mathbf{k}}$ that of the correlated incident mode. (a), (b) Pathways selected by the normal correlations, and (c), (d) pathways selected by the anomalous correlations.}
\label{fig:TA_channels}
\end{figure}

Before investigating possibilities for manipulating the incident field, it is useful to examine the possible arrangements of the states imposed by the Lorentzian profiles in Eq.~(\ref{eq:TA_full}) and the energy conservation conditions in Eq.~(\ref{eq:SF_full}). The two Lorentzians in Eq.~(\ref{eq:TA_full}) become resonant at $\omega=E_l-E_m$ and $\omega=E_n-E_l$, respectively, that is, when the detected photon connects the state $|l\rangle$ to one of the two initially populated states. The two delta functions in Eq.~(\ref{eq:SF_full}), in turn, fix the energy of the incident mode that is correlated with the detected one: $\omega_{\mathbf{k}}=E_m-E_n+\omega$ for the normal correlations $\langle \hat a_{\mathbf{k}'_{\omega},\lambda}^{\dagger} \hat a_{\mathbf{k},\nu} \rangle$, and $\omega_{\mathbf{k}}=E_n-E_m-\omega$ for the anomalous correlations $\langle \hat a_{\mathbf{k}'_{\omega},\lambda}^{\dagger} \hat a_{\mathbf{k},\nu}^{\dagger} \rangle$. Combining these two constraints yields the four possible pathways illustrated in Fig.~\ref{fig:TA_channels}. Panels (a) and (b) correspond to the normal correlations and illustrate scenarios in which the initially populated states $|m\rangle$ and $|n\rangle$ are probed to the upper-lying and lower-lying state $|l\rangle$, respectively. The detected photon is resonant with the transition from one of the initial states to $|l\rangle$, while the correlated incident photon connects either the same state, producing stationary portion of the signal, or a different initial state to the same final state $|l\rangle$. The energies of the detected and the incident photon differ by the splitting of the initially populated states, so that the measurement couples spectral components of the incident field separated by $E_n-E_m$. Panels (c) and (d), demonstrate situations in which the state $|l\rangle$ is located energetically between the initial states $|m\rangle$ and $|n\rangle$. These pathways are selected by the anomalous correlations, for which the detected and the incident photon share the energy splitting between $|m\rangle$ and $|n\rangle$. 

The summation over the wave vector $\mathbf{k}$ in Eq.~(\ref{eq:SF_full}), which arises from considering a general electric field described by Eq.~(\ref{eq:E_I_decomp}), can be simplified by noting that, in a typical transient absorption spectroscopy geometry, the radiation source and detector are positioned along the same propagation axis, with the target located between them. Accordingly, under the idealized assumption of a perfectly collimated incident beam, the only nonzero matrix elements of the beam density matrix $\hat{\rho}_{\mathrm b}(t_i)$ are those which correspond to wave vectors $\mathbf{k}$ that are collinear with the detected wave vector $\mathbf{k}'_{\omega}$. The summation over $\mathbf{k}$ can therefore be replaced by the corresponding sum over frequencies and polarizations for this propagation direction. Therefore, the field-dependent spectral factor can be simplified as
\begin{widetext}
\begin{equation}
\label{eq:SF_simple}
\begin{split}
    \mathcal{P}_{mn}^{ij}(\omega) & \simeq
    \sum_{\lambda,\nu} \Bigg[ 
        \sqrt{E_m-E_n+\omega}
            \left( \epsilon_{\mathbf{k}'_{\omega},\lambda}^{i} \right)^* \epsilon_{\mathbf{k}',\nu}^{j}
            \left\langle \hat a_{\mathbf{k}'_{\omega},\lambda}^{\dagger} \hat a_{\mathbf{k}',\nu} \right\rangle
        - \sqrt{E_n-E_m-\omega}
            \left( \epsilon_{\mathbf{k}'_{\omega},\lambda}^{i} \right)^* (\epsilon_{\mathbf{k}'',\nu}^{j})^*
            \left\langle \hat a_{\mathbf{k}'_{\omega},\lambda}^{\dagger} \hat a^{\dagger}_{\mathbf{k}'',\nu} \right\rangle
    \Bigg],
\end{split}
\end{equation}
\end{widetext}
where the wave vectors $\mathbf{k}'$ and $\mathbf{k}''$ are defined as
\begin{equation}
    \mathbf{k}'=\alpha_{\text{fs}}(E_m-E_n+\omega) \mathbf{k}'_{\omega} / |\mathbf{k}'_{\omega}|,
\end{equation}
and
\begin{equation}
    \mathbf{k}''=\alpha_{\text{fs}}(E_n-E_m-\omega) \mathbf{k}'_{\omega} / |\mathbf{k}'_{\omega}|.
\end{equation}
The obtained expression, Eq.~(\ref{eq:SF_simple}), provides a convenient starting point for investigating how the transient absorption spectrum can be modified by varying the properties of the incident field. In particular, it allows us to explore the effects of quantum correlations between different field modes, as we denonstrate next.

\section{Applications}
\label{sec:applications}

In this section, we consider several examples illustrating how manipulating the properties of light, including purely quantum features such as coherence, can modify the absorption spectrum and potentially suppress noise and enhance the absorption signal.

\subsection{Incoherent versus coherent light}
\label{sec:probe_spectral_coherence}
\begin{figure*}[t]
    \centering
    \includegraphics[width=\textwidth]{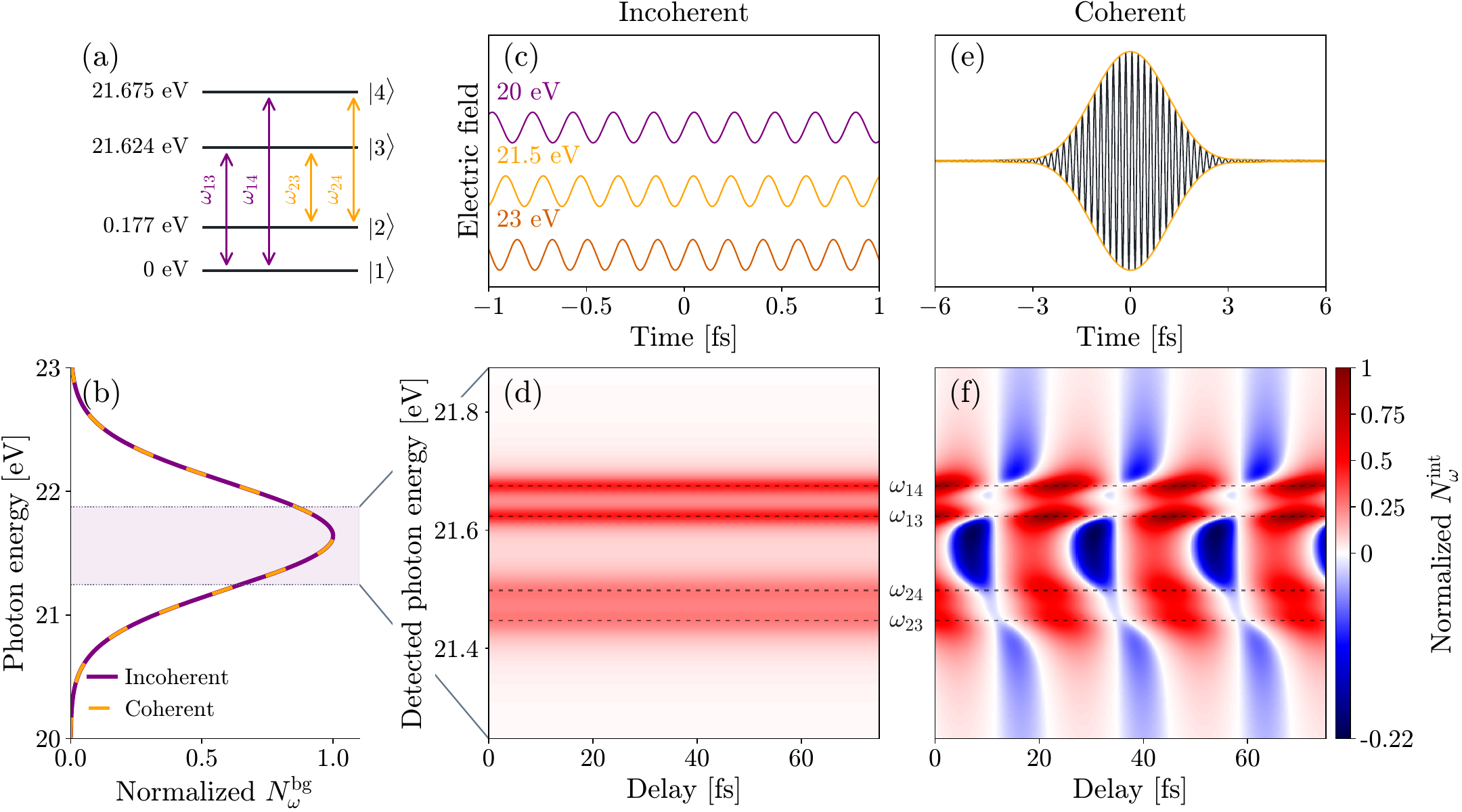}
    \caption{Influence of the spectral coherence of the probe field on the transient absorption signal from a time-evolving target. (a)~Schematic energy-level diagram and the electric dipole couplings between the levels. Initial state is prepared as a coherent superposition of the two lowest states, $\rho_{11}=\rho_{22}=\rho_{12}=\rho_{21}=1/2$. (b) Normalized incident photon-number spectrum $N_{\omega}^{\mathrm{bg}}$, which is taken to be identical for the incoherent and coherent probes. The shaded band marks the detected photon energy range displayed in (d) and (f). (c) Spectrally incoherent probe: representative monochromatic components at $20$, $21.5$, and $23$~eV that carry no mutual phase relation and therefore do not form a collective temporal waveform. (d) Normalized interference contribution $N_{\omega}^{\mathrm{int}}$, reflecting a number of photons absorbed and/or emitted from the target. The delay $\tau$ defines the temporal position of the probe field relative to the target evolution. It is seen that the incoherent probe yields a delay-independent signal despite the nonzero coherence of the target. The gray dashed lines indicate the four transition energies $\omega_{13}$, $\omega_{14}$, $\omega_{23}$, and $\omega_{24}$, shown in (a). (e) Spectrally coherent probe: expectation value of the electric field $\langle\hat{\mathbf E}_{\mathrm I}(t)\rangle$ (black solid line) and its Gaussian envelope (orange solid line). (f) The spectrum obtained with coherent probe field. The beatings of the absorption lines with a period $T=2\pi\hbar/(E_2-E_1)\approx23.4$~fs clearly reflect the coherent dynamics of the target.}
    \label{fig:probe_spectral_coherence}
\end{figure*}

To demonstrate the role of correlations between light modes, we consider the absorption signal of a time-evolving system probed through transitions to a higher-lying excited states. We assume the system is composed of four levels with energies $E_1 = 0$~eV, $E_2 = 0.177$~eV, $E_3 = 21.624$~eV, and $E_4 = 21.675$~eV, coupled via dipole transitions such that $\mathbf{d}_{13} = \mathbf{d}_{23} = \mathbf{d}_{14} = \mathbf{d}_{24}=(1,0,0)$~a.u., along with the corresponding Hermitian conjugates, while all other dipole matrix elements are set to zero. The schematic energy-level structure and the corresponding couplings in the system are shown in Fig.~\ref{fig:probe_spectral_coherence}(a). This arrangement of levels and the couplings between them mimic the spin-orbit split, $3s^2 3p^5 (^2P^{\circ}_{3/2})$ and $3s^2 3p^5 (^2P^{\circ}_{1/2})$, ground electronic state of Ar$^+$ together with the excited $3s^2 3p^4 (^1D)3d (^2P_{3/2})$ and $3s^2 3p^4 (^1D)3d (^2P_{1/2})$ electronic states that we will use for probing the dynamics~\cite{NIST_ASD_2024,chakraborty2026}.

We populate the two lowest states equally, $\rho_{11}=\rho_{22}=1/2$, and assume perfect coherence between them, $\rho_{12}=\rho_{21}=1/2$. All other states are assumed to be unpopulated prior to the arrival of the probe field. The energy splitting between the two lowest states gives rise to coherent oscillations of the electron density with a period $T=2\pi\hbar/(E_2-E_1)\approx23.4$~fs. To account for the finite spectral width of the transitions, we introduce a phenomenological decay rate $\Gamma=0.03$~eV for the final states, corresponding to a lifetime $\hbar/\Gamma\approx22$~fs. Population decay and dephasing of the two lower states are neglected.

We consider the probe field propagating along $z$ direction and take the two orthogonal polarization vectors as $\boldsymbol{\epsilon}_{\mathbf{k},1}=(1,0,0)$ and $\boldsymbol{\epsilon}_{\mathbf{k},2}=(0,1,0)$. We assume the probe is linearly polarized along $x$, such that only the $\lambda=1$ polarization mode is populated. Under these conditions, each field mode is uniquely characterized by its frequency $\omega$. In what follows, we therefore use $\omega$ as the mode index.

The spectral distribution of the photons in the field is assumed to be Gaussian,
\begin{equation}
    f(\omega)=
    \frac{1}{(2\pi\sigma^2)^{1/4}}
    \exp\left[
    -\frac{(\omega-\omega_0)^2}{4\sigma^2}
    \right],
\end{equation}
where $\omega_0$ is the central photon energy and $\sigma$ denotes the spectral bandwidth of the field. Accordingly, the diagonal elements of the field correlation matrix are given by
\begin{equation}
    \left\langle \hat a_{\omega}^{\dagger} \hat a_{\omega} \right\rangle = n |f(\omega)|^2,
\end{equation}
where $n$ denotes the total number of photons in the probe field. Throughout this section, we use $\omega_0=21.5$~eV and $\sigma=1$~eV. Although the absolute photon number does not affect the results presented here, we assume it is sufficiently large to yield a detectable signal above the background, yet low enough to avoid nonlinear interactions with the target. The utilized photon distribution is shown in Fig.~\ref{fig:probe_spectral_coherence}(b).

We first examine the transient absorption spectrum obtained with a spectrally incoherent probe. In this case, different frequency components are mutually unrelated, and the corresponding field correlations are diagonal in frequency, such that 
\begin{equation}
    \left\langle \hat a_{\omega}^{\dagger} \hat a_{\omega'} \right\rangle=0 \qquad \text{for} \qquad \omega \neq \omega'.
\end{equation}
The absence of correlations between different frequency modes prevents the formation of a well-defined temporal waveform. Instead, the applied field represents a combination of individual monochromatic components statistically combined across the utilized frequency range, as schematically illustrated in Fig.~\ref{fig:probe_spectral_coherence}(c). The transient absorption spectrum obtained from this probe is shown in Fig.~\ref{fig:probe_spectral_coherence}(d). As can be seen, the absorption signal is stationary in time despite the presence of nonzero electronic coherence in the target system. The spectrum contains four peaks corresponding to the four allowed transitions between the two initially populated states and the two higher-lying excited states. The absence of spectral coherence in the probe prevents interference between different frequency components, so that the measured spectrum reflects the independent absorption contributions associated with $\omega=\omega'$ transitions.

A coherent probe field, in contrast, maintains well-defined phase relations between its spectral components. The corresponding field correlations are therefore nonzero for different frequencies, with
\begin{equation}
    \left\langle
    \hat a_{\omega}^{\dagger}
    \hat a_{\omega'}
    \right\rangle
    =
    n f(\omega) f(\omega').
\end{equation}
Unlike the incoherent probe, for which the first moments of the field modes vanish, a coherent state has nonzero field amplitudes. Consequently, the expectation value of the electric field operator, $\langle\hat{\mathbf E}_{\mathrm I}(t)\rangle$, shown in Fig.~\ref{fig:probe_spectral_coherence}(e), forms a well-defined temporal waveform whose envelope is determined by the coherent superposition of the corresponding spectral components. The resulting transient absorption spectrum, shown in Fig.~\ref{fig:probe_spectral_coherence}(f), exhibits pronounced temporal oscillations with a period of $\approx23.4$~fs, reflecting the period of the underlying electronic dynamics. These oscillations arise from interference between probe frequency components $\omega$ and $\omega'=\omega+E_m-E_n$, coupled through the evolving electronic coherence of the target. Thus, spectral coherence of the probe enables TR-TAS to capture temporal information about the target dynamics that is inaccessible with a spectrally incoherent field.

\subsection{Enhancing oscillations of spectral lines with squeezed light}
\label{sec:sqp-picture}

\begin{figure*}[t]
    \centering
    \includegraphics[width=\textwidth]{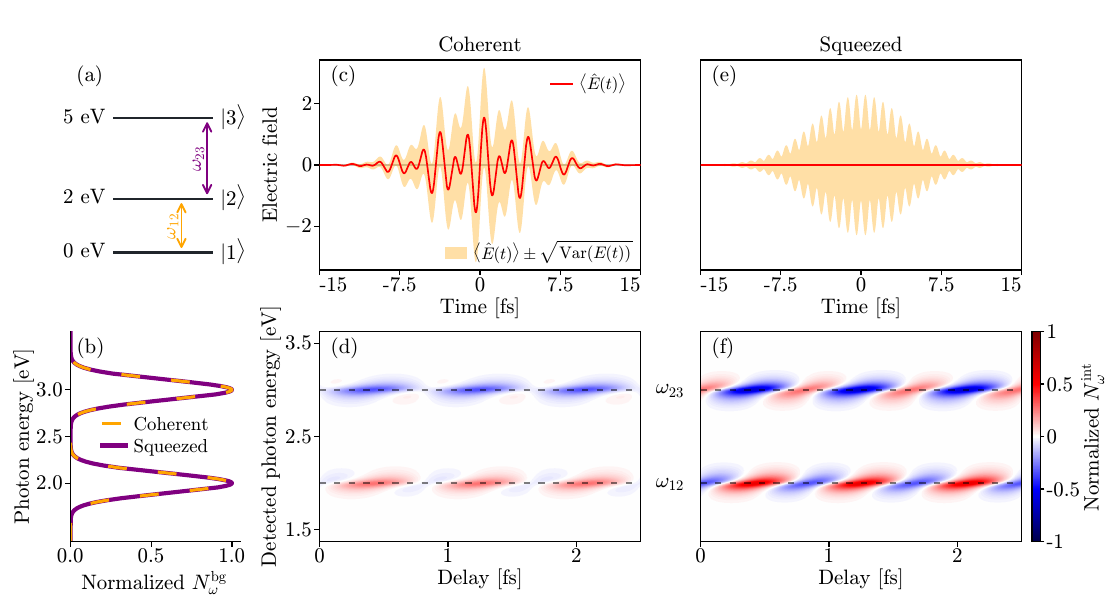}
    \caption{Comparison of a coherent and a squeezed vacuum probe with identical spectral photon-number distributions. (a) Three-level target with the corresponding couplings between the levels. The initial state is prepared as $\rho_{11}=\rho_{33}=\rho_{13}=\rho_{31}=1/2$. (b) Normalized incident photon spectrum $N_{\omega}^{\mathrm{bg}}$, consisting of two bands centered at $\omega_{12}$ and $\omega_{23}$ and identical for both coherent and squeezed probes. (c), (e) Expectation value of the electric field $\langle\hat E(t)\rangle$ (red solid line) together with its variance (shaded area) for the coherent and the squeezed probe. For the squeezed vacuum the mean field vanishes, and only the variance carries the temporal structure of the probe. (d), (f) Normalized interference contribution $N_{\omega}^{\mathrm{int}}(\tau)$ on a common color scale. Both probes produce absorption lines that oscillate with the period $T=2\pi\hbar/(E_3-E_1)\approx0.83$~fs, set by the target coherence. However, the squeezed probe yields larger oscillation amplitudes, providing greater contrast and making the absorption lines more clearly distinguishable from the background.}
    \label{fig:sqb-oscillation-advantage}
\end{figure*}

As we have demonstrated in the previous section, coherences between field modes play a crucial role in the ability to resolve quantum dynamics with TR-TAS. Beyond the possibility to switch intermode coherences on and off, the emergence of experimental techniques capable of generating intense ultrashort quantum states of light~\cite{gorlach2020,gorlach2023,rasputnyi2024,sennary2025light,kern2026}, including squeezed light, opens new avenues for manipulating light coherences in previously inaccessible ways. Here, we demonstrate how the magnitude of light coherences can be engineered to enhance oscillations of spectral lines beyond the limits attainable with coherent light.

We consider linearly polarized light propagating along a fixed direction, such that each mode is uniquely characterized by its frequency $\omega$, as in the previous section. Before turning to a concrete example, we examine the correlations between two field modes with frequencies $\omega$ and $\omega' \neq \omega$ that contribute to the transient absorption signal. To facilitate a direct comparison between coherent and squeezed light, we impose the condition that the mean photon number in each frequency mode is identical for the two states, such that
\begin{equation}
\label{eq:photon_number}
    \langle \hat a^\dagger_{\omega}\hat a_{\omega}\rangle_{\text{coh}/\text{sq}} =
    \langle \hat a^\dagger_{\omega'}\hat a_{\omega'}\rangle_{\text{coh}/\text{sq}} \\ 
    = n.
\end{equation}
This condition ensures that any differences in the transient absorption spectra arise from the distinct quantum correlations of the probe field rather than from differences in the spectral photon-number distribution.

To analyze the bounds within which the field coherences can be manipulated, we make use of the Cauchy--Schwarz inequality, which for two arbitrary operators $\hat A$ and $\hat B$ states
\begin{equation}
\label{eq:CS_ineq}
    \left|\langle \hat A^\dagger \hat B\rangle\right|
    \leq
    \sqrt{
        \langle \hat A^\dagger \hat A\rangle
        \langle \hat B^\dagger \hat B\rangle
    }.
\end{equation}
Choosing $\hat A=\hat a_{\omega}$ and $\hat B=\hat a_{\omega'}$ gives the bound on the normal intermode coherence,
\begin{equation}
\label{eq:CS_ineq_normal}
    \left|
        \left\langle
        \hat a^\dagger_{\omega}\hat a_{\omega'}
        \right\rangle
    \right|
    \leq
    \sqrt{
        \left\langle
            \hat a^\dagger_{\omega}\hat a_{\omega}
        \right\rangle
        \left\langle
            \hat a^\dagger_{\omega'}\hat a_{\omega'}
        \right\rangle
    } = n.
\end{equation}
On the other hand, choosing $\hat A=\hat a_{\omega}$ and $\hat B=\hat a^\dagger_{\omega'}$ yields a bound on the anomalous correlation,
\begin{equation}
\label{eq:CS_ineq_anom}
    \left|
        \left\langle
            \hat a^\dagger_{\omega}\hat a^\dagger_{\omega'}
        \right\rangle
    \right|
    \leq
    \sqrt{
        \left\langle
            \hat a^\dagger_{\omega}\hat a_{\omega}
        \right\rangle
        \left\langle
            \hat a_{\omega'}\hat a^\dagger_{\omega'}
        \right\rangle
    }
    =
    \sqrt{
        n(n+1)
    },
\end{equation}
where we have used the bosonic commutation relation $\hat a_{\omega'}\hat a^\dagger_{\omega'} = 1 + \hat a^\dagger_{\omega'}\hat a_{\omega'}$ to compute the required expectation value.

Considering a multimode coherent state of light, $|\{\alpha\}\rangle$, for which
\begin{equation}
\label{eq:coherent_w_wp}
    \hat a_{\omega} |\{\alpha\}\rangle = \alpha_{\omega} |\{\alpha\}\rangle
    \quad \text{and} \quad
    \hat a_{\omega'} |\{\alpha\}\rangle = \alpha_{\omega'} |\{\alpha\}\rangle,
\end{equation}
the normal intermode coherence is
\begin{equation}
    \left\langle
        \hat a^\dagger_{\omega}\hat a_{\omega'}
    \right\rangle_{\text{coh}} = \alpha^{*}_{\omega} \alpha_{\omega'}.
\end{equation}
Using the condition $|\alpha_{\omega}|^2 = |\alpha_{\omega'}|^2 = n$, imposed by Eq.~(\ref{eq:photon_number}), we obtain
\begin{equation}
    \left| 
    \left\langle
        \hat a^\dagger_{\omega}\hat a_{\omega'}
    \right\rangle_{\text{coh}} 
    \right| = \left| \alpha^{*}_{\omega} \right| \left| \alpha_{\omega'} \right| = n.
\end{equation}
Comparing this result with Eq.~(\ref{eq:CS_ineq_normal}), we see that coherent light saturates the Cauchy--Schwarz bound. Therefore, for fixed mean photon populations, the magnitude of the normal intermode coherence cannot be increased beyond its coherent-state value. Consequently, the amplitude of the absorption lines cannot be enhanced beyond the coherent light limit by increasing the normal intermode coherence alone.

For the anomalous correlation, the same coherent light gives
\begin{equation}
\label{eq:anomalous_correlations_coh}
    \left| 
    \left\langle
        \hat a^\dagger_{\omega}\hat a^\dagger_{\omega'}
    \right\rangle_{\text{coh}} 
    \right| = \left| \alpha^{*}_{\omega} \right| \left| \alpha^{*}_{\omega'} \right| = n.
\end{equation}
This value is smaller than the corresponding Cauchy--Schwarz bound, Eq.~(\ref{eq:CS_ineq_anom}), for any finite $n$. Thus, the anomalous field correlation provides a route for designing quantum states of light that could increase the contrast of oscillations in the transient absorption spectrum toward its quantum-mechanical upper bound, potentially giving access to stronger temporal modulations of spectral features than those achievable with conventional coherent light.

As can be seen from Eqs.~(\ref{eq:TA_full}) and~(\ref{eq:SF_full}), anomalous field correlations play an essential role in the transient absorption spectrum for systems with the level structures shown in Figs.~\ref{fig:TA_channels}(c) and~(d). Therefore, we consider a three-level system with energies $E_1=0$~eV, $E_2=2$~eV, and $E_3=5$~eV, populated such that $\rho_{11}=\rho_{33}=\rho_{13}=\rho_{31}=1/2$. The levels are coupled via dipole transitions such that $\mathbf{d}_{12}=\mathbf{d}_{23}=(1,0,0)$~a.u., along with the corresponding Hermitian conjugates, while all other dipole matrix elements are set to zero. The schematic energy-level structure and the corresponding couplings in the system are shown in Fig.~\ref{fig:sqb-oscillation-advantage}(a). We assume a decay rate $\Gamma=0.05$~eV for the intermediate state and neglect all other population decay and dephasing mechanisms.

We examine a probe field composed of photons distributed over two spectral domains, $A$ and $B$, with Gaussian spectral distributions $f_A(\omega)$ and $f_B(\omega)$, centered at $\omega_0^A=E_2-E_1$ and $\omega_0^B=E_3-E_2$, respectively. Both distributions have the same spectral bandwidth, $\sigma=0.25$~eV. The utilized photon distributions are shown in Fig.~\ref{fig:sqb-oscillation-advantage}(b). Such a two-band probe field can, for example, represent two neighboring harmonics of a HHG spectrum.

To describe the two spectral bands as independent field modes, we introduce the collective annihilation operators
\begin{equation}
    \hat a_A=\int d\omega\, f^{*}_A(\omega)\hat a_\omega
    \quad \text{and} \quad
    \hat a_B=\int d\omega\, f^{*}_B(\omega)\hat a_\omega,
\end{equation}
for which we assume negligible overlap, $f_A(\omega)f_B(\omega) \simeq 0$, so that the collective operators obey the canonical bosonic commutation relations $[\hat a_I,\hat a_J^\dagger]=\delta_{IJ}$, where $I,J \in \{A,B\}$. Each operator therefore describes a single bosonic mode whose spectral profile is given by the corresponding Gaussian distribution, rather than a single frequency component. In this representation, the probe field is described by two orthogonal collective modes associated with the spectral domains $A$ and $B$. The coherent-light correlations are then fully specified by Eqs.~(\ref{eq:coherent_w_wp})--(\ref{eq:anomalous_correlations_coh}), allowing us to focus directly on the corresponding correlations for squeezed light.

As an illustrative example, we consider a squeezed vacuum state,
\begin{equation}
\label{eq:multimode_squeezed_state}
    |\Psi\rangle =
    \hat{S}(\boldsymbol{\beta})|0\rangle,
\end{equation}
where $|0\rangle$ denotes the electromagnetic vacuum state and $\hat{S}(\boldsymbol{\beta})$ is the multimode squeezing operator (see Appendix~\ref{app:squeezing}),
\begin{equation}
\label{eq:multimode_squeezing_operator}
    \hat{S}(\boldsymbol{\beta})
    =
    \exp\left\{
        \frac{1}{2}
        \left(
            \hat{\mathbf{a}}^{\dagger}
                \boldsymbol{\beta}
            \hat{\mathbf{a}}^{\dagger T}
                -
            \hat{\mathbf{a}}^{\text{T}}
                \boldsymbol{\beta}^{*}
            \hat{\mathbf{a}}
        \right)
    \right\}.
\end{equation}
Here, $\hat{\mathbf{a}}$ is the column vector of annihilation operators associated with the optical modes and $\boldsymbol{\beta}$ is a complex symmetric matrix specifying the squeezing. In the basis of $\hat{a}_A$ and $\hat{a}_B$ operators, we define the column vector of annihilation operators as $\hat{\mathbf{a}}=(\hat{a}_A,\hat{a}_B)^T$, and choose the squeezing matrix as
\begin{equation}
    \boldsymbol{\beta}
    =
    R
    \begin{pmatrix}
        0 & 1\\
        1 & 0
    \end{pmatrix},
\end{equation}
where $R\geq 0$ is the squeezing parameter. This choice generates two-mode squeezing, creating correlated photon pairs across the two spectral domains.

The utilized state has equal mean photon populations in the two collective modes (see Appendix~\ref{app:squeezing} for a derivation of the general relations for a squeezed light),
\begin{equation}
\label{eq:squeezed_n}
    \langle\hat{a}_A^\dagger\hat{a}_A\rangle_{\text{sq}}
    =
    \langle\hat{a}_B^\dagger\hat{a}_B\rangle_{\text{sq}}
    =
    \sinh^2 R = n,
\end{equation}
and we choose the squeezing parameter $R$ such that the mean total photon number matches that of the coherent probe considered above. In the chosen squeezing regime, the interband normal correlation vanishes, $\langle\hat{a}_A^\dagger\hat{a}_B\rangle_{\text{sq}}=0$, while the anomalous correlation becomes
\begin{equation}
    \langle\hat{a}_A^\dagger\hat{a}_B^\dagger\rangle_{\text{sq}}
    =
    \sinh(R) \cosh(R) = \sqrt{n(n+1)},
\end{equation}
where we used Eq.~(\ref{eq:squeezed_n}) and the identity $\cosh^2(R)=1+\sinh^2(R)$. Therefore, the chosen squeezing state saturates the Cauchy--Schwarz bound, Eq.~(\ref{eq:CS_ineq_anom}), for the anomalous correlation between modes $A$ and $B$.

\begin{figure}[t]
    \centering
    \includegraphics[width=\columnwidth]{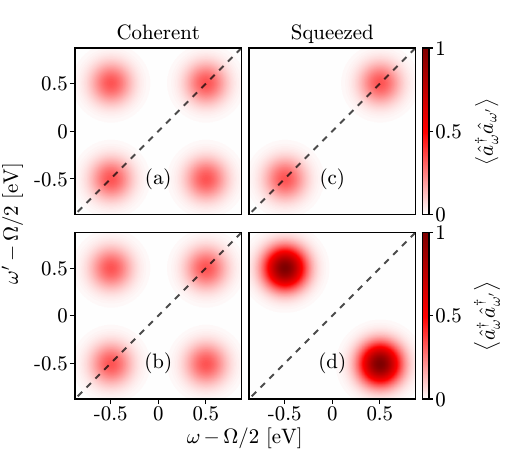}
    \caption{Normal, $\langle\hat a^\dagger_{\omega}\hat a_{\omega'}\rangle$, and anomalous, $\langle\hat a^\dagger_{\omega}\hat a^\dagger_{\omega'}\rangle$, field correlations of the coherent [(a),(b)] and two-mode squeezed vacuum [(c),(d)] probes with identical spectral photon-number distributions. The energy axes are depicted  relative to $\Omega/2=(E_3-E_1)/2=2.5$~eV, and all panels share a common color scale. The dashed line marks $\omega=\omega'$. For the coherent probe both types of correlations contain contributions within and between the two spectral bands, whereas for the squeezed vacuum the normal correlations remain within each band and the anomalous correlations couple the bands with a larger magnitude than in the coherent case.}
    \label{fig:sqb-photon-correlations}
\end{figure}

Switching back to the basis of frequency modes, the normal correlations can be expressed as
\begin{equation}
    \left\langle
        \hat a^\dagger_{\omega}\hat a_{\omega'}
    \right\rangle = 
    \sum_{I,J \in \{A,B\}} f^*_I (\omega) f_J (\omega') 
    \langle\hat{a}_I^\dagger\hat{a}_J\rangle,
\end{equation}
and, similarly, the anomalous correlations are given by
\begin{equation}
    \left\langle
        \hat a^\dagger_{\omega}\hat a^\dagger_{\omega'}
    \right\rangle
    =
    \sum_{I,J \in \{A,B\}} f^*_I(\omega) f^*_J(\omega')
    \left\langle\hat{a}_I^\dagger\hat{a}_J^\dagger\right\rangle.
\end{equation}
Consequently, for the coherent probe with zero relative phase, the normal and anomalous correlations are
\begin{equation}
\label{eq:field_corr_coh}
\begin{aligned}
    \left\langle
        \hat a^\dagger_{\omega}\hat a_{\omega'}
    \right\rangle_{\mathrm{coh}}
    &=
    n\big[f_A^*(\omega)+f_B^*(\omega)\big]
    \big[f_A(\omega')+f_B(\omega')\big],
    \\
    \left\langle
        \hat a^\dagger_{\omega}\hat a^\dagger_{\omega'}
    \right\rangle_{\mathrm{coh}}
    &=
    n\big[f_A^*(\omega)+f_B^*(\omega)\big]
    \big[f_A^*(\omega')+f_B^*(\omega')\big].
\end{aligned}
\end{equation}
Thus, for coherent light, both normal and anomalous correlations contain contributions within and between the two spectral bands. These correlations are depicted in panels~(a) and~(b), respectively, of Fig.~\ref{fig:sqb-photon-correlations}. In contrast, for the two-mode squeezed vacuum, the corresponding correlations are
\begin{equation}
\label{eq:field_corr_sq}
\begin{aligned}
    \left\langle
        \hat a^\dagger_{\omega}\hat a_{\omega'}
    \right\rangle_{\mathrm{sq}}
    &=
    n\left[
        f_A^*(\omega)f_A(\omega')
        +f_B^*(\omega)f_B(\omega')
    \right],
    \\
    \left\langle
        \hat a^\dagger_{\omega}\hat a^\dagger_{\omega'}
    \right\rangle_{\mathrm{sq}}
    &=
    \sqrt{n(n+1)}
    \left[
        f_A^*(\omega)f_B^*(\omega')
        +f_B^*(\omega)f_A^*(\omega')
    \right],
\end{aligned}
\end{equation}
which are depicted in panels~(c) and~(d) of Fig.~\ref{fig:sqb-photon-correlations}. As can be seen, for the considered squeezed-vacuum state, the normal correlations involve contributions only within each band, whereas the anomalous correlations couple the two bands. Importantly, the anomalous coherences between bands $A$ and $B$ are larger in magnitude than those for coherent light. This enhancement suggests that the squeezed light might produce more pronounced oscillations of the absorption lines than a coherent probe, as we demonstrate next.

The coherent electric field, depicted by the expectation value of the electric field operator, $\langle\hat{\mathbf E}_{\mathrm I}(t)\rangle$, in Fig.~\ref{fig:sqb-oscillation-advantage}(c), is simply a combination of Gaussian laser pulses with energy profiles shown in Fig.~\ref{fig:sqb-oscillation-advantage}(b). The resulting temporal profile reflects constructive and destructive interference between the two frequency components, which produces the characteristic modulation of the field amplitude. Figure~\ref{fig:sqb-oscillation-advantage}(c) also shows the variance of the electric field operator, which quantifies fluctuations about its mean value. For the coherent state, these fluctuations follow the temporal envelope of the field, reflecting the time dependence of the mean photon flux. The transient absorption spectrum obtained from this probe is shown in Fig.~\ref{fig:sqb-oscillation-advantage}(d). As expected, the absorption lines exhibit oscillations with period $T\approx0.83$~fs, determined by the energy separation between the coherently populated target states $|1\rangle$ and $|3\rangle$. These oscillations occur out of phase with respect to each other, which directly follows from opposite signs of the corresponding Lorentzian profiles describing transitions up from the state $|1\rangle$ and down from state $|3\rangle$, respectively. Furthermore, these oscillations appear on top of the static background signal arising from the diagonal portion of the anomalous correlations present in coherent state (see Fig.~\ref{fig:sqb-photon-correlations}(b)).

In case of the squeezed vacuum state, the mean field amplitudes $\langle\hat{a}_A\rangle_{\text{sq}}$ and $\langle\hat{a}_B\rangle_{\text{sq}}$ vanish, and, consequently, the expectation value of the electric field operator also vanishes, $\langle\hat{\mathbf E}_{\mathrm I}(t)\rangle_{\text{sq}}=0$. The field variance, however, provides a means of visualizing the temporal structure of the probe, as demonstrated in Fig.~\ref{fig:sqb-oscillation-advantage}(e). Because the chosen squeezed vacuum state has no first-order coherence between modes $A$ and $B$ (see Fig.~\ref{fig:sqb-photon-correlations}(c)), the variance reflect the individual temporal profiles of the two modes without the interference modulation characteristic of the coherent field. The transient absorption spectrum shown in Fig.~\ref{fig:sqb-oscillation-advantage}(f) exhibits two absorption lines that oscillate out of phase with each other, with the same period as those obtained with the coherent probe. Notably, the oscillations have a larger amplitude, demonstrating the enhanced sensitivity of the squeezed-vacuum probe to the target dynamics. 

The changes in the transient absorption spectrum obtained with squeezed light compared with the coherent probe arise directly from the modified field correlation functions [see Eqs.~(\ref{eq:field_corr_coh}) and~(\ref{eq:field_corr_sq}), and Fig.~\ref{fig:sqb-photon-correlations}]. The ability to manipulate photon correlations while keeping the spectral distribution unchanged provides a means of controlling the transient absorption signal through the quantum properties of the probe. These results demonstrate that tailoring quantum correlations in the probe field opens new avenues for enhancing the applicability and the sensitivity of the TR-TAS in experiments probing ultrafast quantum dynamics in matter.

\section{Conclusions}
\label{sec:conclusions}

In conclusion, we report in this paper a fully quantum theory of TR-TAS, in which both the target and the probe field are treated quantum mechanically. Starting from a general framework applicable to a broad range of spectroscopic experiments, we developed the formalism and working equations giving access to the number of photons arriving at the detector at a given photon energy, the directly measurable quantity in photon-counting spectroscopy. We showed that this observable naturally separates into three contributions with distinct physical origins and, to some extent, distinct detection geometries. Among these contributions, we examined the term describing interference between the freely propagating incident field and the field radiated by the target. We demonstrated the importance of this contribution for transmission-based measurements, including absorption and stimulated-emission spectroscopies.

Employing the dipole approximation and evaluating the interference contribution to linear order in the probe field, we obtained a practical expression for modeling the transient absorption spectra of a target prepared in a general nonequilibrium state. The resulting expression closely resembles the familiar semiclassical result in its description of the target's properties, but also contains a previously unexplored contribution that encodes the quantum state of the probe field. This field-dependent spectral factor contains the correlations between the field modes and enforces the required energy conservation conditions, thereby linking the detected photon energy to specific spectral components of the incident field involved in the corresponding interference pathways. By explicitly connecting the energy levels of the target to the photons of the incident field, we examined how different arrangements of the target's energy levels relate to the quantum correlations between field modes. We identified previously unexplored pathways involving anomalous field correlations that have no classical counterparts.

We applied the developed formalism to simulate time-resolved transient absorption spectra of a target prepared in a coherent superposition of states. We demonstrated how coherence between field modes manifests itself in the absorption signal and established a computational procedure for simulating transient absorption in arbitrary targets interacting with arbitrary quantum probe fields. Although it was expected that incoherent light cannot be used to measure the time evolution of a target, no practical equations were available to explicitly simulate such measurements. Accordingly, less extreme scenarios involving partially coherent light or, conversely, light with extreme quantum correlations can be explored straightforwardly using the developed formalism.

To illustrate the capabilities of the developed approach in the regime of strong quantum correlations, we investigated transient absorption driven by squeezed light. We explored the extent to which intermode correlations can be manipulated while keeping the photon-number distributions fixed. Comparing the capabilities of coherent light and the squeezed vacuum state, we found that the latter enables enhanced oscillations in the transient absorption signal through its stronger anomalous intermode correlations. These correlations, however, become relevant only for specific arrangements of the initial target states relative to the energy levels involved in the absorption and emission transitions. Interestingly, the level structure considered here allows these oscillations to be resolved using probe pulses whose temporal profiles extend well beyond the characteristic timescale of the target's coherent dynamics. This finding opens new avenues for attosecond science by suggesting that quantum correlations in the probe field can help access ultrafast dynamics without requiring the probe pulse itself to be comparably short.

This work represents a proof-of-concept theoretical study, offering a foundation for future theoretical advancements and experimental investigations. The two-band probe field used in our example can, for instance, be realized with two neighboring harmonics of a high-harmonic generation spectrum, and the rapid development of sources of intense ultrashort quantum light makes the required states increasingly accessible. We leave the detailed discussion of possible extensions for future research studies. Among them, the polarization degrees of freedom retained in our expressions allow correlations between polarization components of the probe to be engineered as well, which may offer a route to shaping individual spectral lines. Nonlinear and background-free signals, important for modeling HHG process and various scattering spectroscopy techniques, are contained in the radiation contribution and can be analyzed within the same formalism by going beyond the linear order in the probe field. Finally, the present analysis concerns the mean detected photon number, while the fluctuations of the signal, and therefore the precision with which the target dynamics can be extracted from a given number of probe photons, constitute a separate question that the developed quantum description is well suited to address. We hope that our work will motivate further studies exploring the use of tailored quantum light for probing ultrafast quantum dynamics in matter.

\begin{acknowledgments}
The computational part of this research is based upon High Performance Computing (HPC) resources supported by the University of Arizona TRIF, UITS, and Research, Innovation, and Impact (RII) and maintained by the UArizona Research Technologies department.
\end{acknowledgments}

\section*{Data availability}
The data that supports the findings of this study are available within the article.

\appendix

\begin{widetext}
    
\section{Time evolution of the field operators}
\label{app:field_opp}
We derive here the time evolution of the annihilation and creation operators of the quantized electromagnetic field. As we demonstrated in Eq.~(\ref{eq:Nw}), the field operators acquire time dependence in the Heisenberg picture according to
\begin{equation}
\label{eq:a_heisen}
    \hat{a}_{\mathbf{k},\lambda}(t) =
    \hat{U}^{\dagger}(t,t_i)
    \hat{a}_{\mathbf{k},\lambda}
    \hat{U}(t,t_i),
\end{equation}
where $\hat{U}(t,t_i)$ is the time-evolution operator generated by the full Hamiltonian. The Heisenberg equation of motion then reads
\begin{equation}
\label{eq:heisen_eq}
    i\frac{d}{dt}
    \hat{a}_{\mathbf{k},\lambda}(t)
    =
    \left[
        \hat{a}_{\mathbf{k},\lambda}(t),
        \hat{H}
    \right] =
    \hat{U}^{\dagger}(t,t_i)
    \left[
        \hat{a}_{\mathbf{k},\lambda},
        \hat{H}
    \right]
    \hat{U}(t,t_i).
\end{equation}
Here, we have used the fact that, for a time-independent Hamiltonian, $[\hat{H},\hat{U}(t,t_i)]=0$, and hence $\hat{H}=\hat{U}^{\dagger}(t,t_i)\hat{H}\hat{U}(t,t_i)$. Together with the unitarity condition $\hat{U}^{\dagger}(t,t_i)\hat{U}(t,t_i)=\hat{1}$ and Eq.~(\ref{eq:a_heisen}), this allows the commutator to be expressed as a unitary transformation of the corresponding Schr\"odinger-picture commutator.

Decomposing the total Hamiltonian using Eqs.~(\ref{eq:H_total}) and~(\ref{eq:H_0}), we can write
\begin{equation}
\label{eq:heisen_a}
    \left[
        \hat{a}_{\mathbf{k},\lambda},
        \hat{H}
    \right] =
    \left[
        \hat{a}_{\mathbf{k},\lambda},
        \hat{H}_{\text{b}}
    \right]
    +\left[
        \hat{a}_{\mathbf{k},\lambda},
        \hat{\mathcal{V}}
    \right],
\end{equation}
where we have used the fact that the target Hamiltonian, $\hat{H}_{\text{t}}$, does not contribute to Eq.~(\ref{eq:heisen_a}) because it acts exclusively on the target degrees of freedom and therefore commutes with the field annihilation operator. Using the free-field Hamiltonian, Eq.~(\ref{eq:Hb}), and the bosonic commutation relations, the first commutator gives 
\begin{equation}
    \left[
        \hat{a}_{\mathbf{k},\lambda},
        \hat{H}_{\mathrm{b}}
    \right]
    =
    \omega_{\mathbf{k}}
    \hat{a}_{\mathbf{k},\lambda}.
\end{equation}
The interaction contribution for the minimal-coupling Hamiltonian, Eq.~(\ref{eq:H_int}), gives
\begin{equation}
\label{eq:a_Hint_comm_initial}
    \left[
        \hat{a}_{\mathbf{k},\lambda},
        \hat{\mathcal{V}}
    \right]
    =
    -\alpha_{\mathrm{fs}}
    \int d^3r\,
    \hat{\mathbf{j}}(\mathbf{r})
    \cdot
    \left[
        \hat{a}_{\mathbf{k},\lambda},
        \hat{\mathbf{A}}(\mathbf{r})
    \right]
    +
    \frac{\alpha_{\mathrm{fs}}^2}{2}
    \int d^3r\,
    \hat{n}(\mathbf{r})
    \left[
        \hat{a}_{\mathbf{k},\lambda},
        \hat{\mathbf{A}}^2(\mathbf{r})
    \right].
\end{equation}
Using the expansion for the quantized vector potential, Eq.~(\ref{eq:A_op}), and applying bosonic commutation relations yields
\begin{equation}
    \left[
        \hat{a}_{\mathbf{k},\lambda},
        \hat{\mathbf{A}}(\mathbf{r})
    \right]
    =
    \sqrt{
        \frac{2\pi}
        {V\omega_{\mathbf{k}}\alpha_{\mathrm{fs}}^2}
    }
    \boldsymbol{\epsilon}_{\mathbf{k},\lambda}^{*}
    e^{-i\mathbf{k}\cdot\mathbf{r}}.
\end{equation}
The second term in Eq.~(\ref{eq:a_Hint_comm_initial}) can be transformed as
\begin{equation}
    \left[
        \hat{a}_{\mathbf{k},\lambda},
        \hat{\mathbf{A}}^2(\mathbf{r})
    \right] =
    \left[
        \hat{a}_{\mathbf{k},\lambda},
        \hat{\mathbf{A}}(\mathbf{r})
    \right]
    \cdot
    \hat{\mathbf{A}}(\mathbf{r})
    + \hat{\mathbf{A}}(\mathbf{r}) \cdot
    \left[
        \hat{a}_{\mathbf{k},\lambda},
        \hat{\mathbf{A}}(\mathbf{r})
    \right]
    =
    2 \hat{\mathbf{A}}(\mathbf{r}) \cdot
    \left[
        \hat{a}_{\mathbf{k},\lambda},
        \hat{\mathbf{A}}(\mathbf{r})
    \right],
\end{equation}
where we applied the product rule for commutators and utilized the fact that $\left[ \hat{a}_{\mathbf{k},\lambda}, \hat{\mathbf{A}}(\mathbf{r}) \right]$ is a number and thus commutes with $\hat{\mathbf{A}}(\mathbf{r})$. Combining terms together, we can write
\begin{equation}
    \left[
        \hat{a}_{\mathbf{k},\lambda},
        \hat{\mathcal{V}}
    \right] =
    - \sqrt{\frac{2\pi}{V\omega_{\mathbf{k}}}}
    \int d^3r\,
    \boldsymbol{\epsilon}_{\mathbf{k},\lambda}^{*}
    \cdot
    \hat{\mathbf{J}}(\mathbf{r})
    e^{-i\mathbf{k}\cdot\mathbf{r}},
\end{equation}
where we have introduced the total current density operator 
\begin{equation}
    \hat{\mathbf{J}}(\mathbf{r}) =
    \hat{\mathbf{j}}(\mathbf{r}) -
    \alpha_{\mathrm{fs}}
    \hat{\mathbf{A}}(\mathbf{r})
    \hat{n}(\mathbf{r}).
\end{equation}

Consequently, the equation of motion for the annihilation operator, Eq.~(\ref{eq:heisen_eq}), takes the form
\begin{equation}
\label{eq:a_heisen_resolved}
    i\frac{d}{dt}
    \hat{a}_{\mathbf{k},\lambda}(t)
    =
    \omega_{\mathbf{k}}
    \hat{a}_{\mathbf{k},\lambda}(t)
    -
    \sqrt{\frac{2\pi}{V\omega_{\mathbf{k}}}}
    \int d^3r\,
    \boldsymbol{\epsilon}_{\mathbf{k},\lambda}^{*}
    \cdot
    \hat{\mathbf{J}}(\mathbf{r},t)
    e^{-i\mathbf{k}\cdot\mathbf{r}},
\end{equation}
where we have applied the $\hat{U}^{\dagger}(t,t_i)$ and $\hat{U}(t,t_i)$ operators thus producing $\hat{a}_{\mathbf{k},\lambda}(t)$ and time-dependent current, $\hat{\mathbf{J}}(\mathbf{r},t)$, terms. Equation~(\ref{eq:a_heisen_resolved}) is a first-order inhomogeneous differential equation. Introducing the integrating factor $e^{i\omega_{\mathbf{k}}t}$, we can rewrite it as
\begin{equation}
    \frac{d}{dt}
    \left[
        e^{i\omega_{\mathbf{k}}t}
        \hat{a}_{\mathbf{k},\lambda}(t)
    \right]
    =
    i\sqrt{\frac{2\pi}{V\omega_{\mathbf{k}}}}
    e^{i\omega_{\mathbf{k}}t}
    \int d^3r\,
    \boldsymbol{\epsilon}_{\mathbf{k},\lambda}^{*}
    \cdot
    \hat{\mathbf{J}}(\mathbf{r},t)
    e^{-i\mathbf{k}\cdot\mathbf{r}}.
\end{equation}
Integrating from the initial time $t_i$ to $t$ gives the formal solution
\begin{equation}
    \hat{a}_{\mathbf{k},\lambda}(t)
    =
    e^{-i\omega_{\mathbf{k}}(t-t_i)}
    \hat{a}_{\mathbf{k},\lambda}
    + i\sqrt{\frac{2\pi}{V\omega_{\mathbf{k}}}}
    \int_{t_i}^{t}dt'\,
    e^{-i\omega_{\mathbf{k}}(t-t')}
    \int d^3r\,
    \boldsymbol{\epsilon}_{\mathbf{k},\lambda}^{*}
    \cdot
    \hat{\mathbf{J}}(\mathbf{r},t')
    e^{-i\mathbf{k}\cdot\mathbf{r}}.
\end{equation}
The first term describes the free evolution of the initial field mode, while the second term represents the field generated by the target current. Taking the Hermitian conjugate gives the corresponding expression for the creation operator,
\begin{equation}
\label{eq:a_decomp_conj}
    \hat{a}_{\mathbf{k},\lambda}^{\dagger}(t)
    =
    e^{i\omega_{\mathbf{k}}(t-t_i)}
    \hat{a}_{\mathbf{k},\lambda}^{\dagger}
    -
    i\sqrt{\frac{2\pi}{V\omega_{\mathbf{k}}}}
    \int_{t_i}^{t}dt'\,
    e^{i\omega_{\mathbf{k}}(t-t')}
    \int d^3r\,
    \boldsymbol{\epsilon}_{\mathbf{k},\lambda}
    \cdot
    \hat{\mathbf{J}}(\mathbf{r},t')
    e^{i\mathbf{k}\cdot\mathbf{r}},
\end{equation}
where we have used the Hermiticity of the current density operator, $\hat{\mathbf{J}}^\dagger(\mathbf{r},t) = \hat{\mathbf{J}}(\mathbf{r},t)$.

\section{Dipole form of the radiative field}
\label{app:dip_field_opp}
We recast here the spatially resolved target current in terms of the time-dependent electric dipole operator. In the long-wavelength limit, $e^{-i\mathbf{k}'_{\omega}\cdot\mathbf{r}} \simeq 1$, the spatially integrated physical current is related to the time derivative of the electric dipole operator according to (see, e.g., Ref.~\cite{milton2006})
\begin{equation}
    \int d^3r\,
        \hat{\mathbf{J}}(\mathbf{r},t) = \frac{d \hat{\mathbf{d}}(t)}{d t},
\end{equation}
where $\hat{\mathbf{d}}(t)=-\int d^3 r\, \mathbf{r}\, \hat{n}(\mathbf{r},t)$ is the Heisenberg-picture electric dipole operator. Consequently, the radiative part of the annihilation operator, Eq.~(\ref{eq:a_rad}), can be transformed as
\begin{equation}
\label{eq:ar_t_dip_approx}
\begin{split}
    \hat{a}^{\text{r}}_{\mathbf{k},\lambda}(t) & = 
        i\sqrt{\frac{2\pi}{V\omega_{\bk}}}
        \int_{t_i}^{t} dt'\,
        e^{-i\omega_{\bk}(t-t')}
        \int d^3r\,
        \boldsymbol{\epsilon}_{\mathbf{k},\lambda}^{*}
        \cdot
        \hat{\mathbf{J}}(\mathbf{r},t')
        e^{-i\mathbf{k}'_{\omega}\cdot\mathbf{r}} \\
    & \simeq
        \sqrt{\frac{2\pi \omega_{\bk}}{V}}
        \int_{t_i}^{t} dt'\, e^{-i\omega_{\bk}(t-t')} 
        \boldsymbol{\epsilon}_{\mathbf{k},\lambda}^{*}
            \cdot
        \hat{\mathbf d}(t').
\end{split}
\end{equation}
Here, we interpret the time integral as the Fourier transform of the radiating current, consistent with the standard frequency-domain relation between the current and dipole, $\tilde{\mathbf{J}}(\omega)=-i\omega\tilde{\mathbf{d}}(\omega)$. In the asymptotic limit $t_i\to-\infty$ and $t_f\to+\infty$, this Fourier transform determines the distribution of emitted photons over frequency and isolates the components of the target dynamics resonant with the detected photon frequency $\omega$. The boundary terms generated by integration by parts correspond to contributions at the temporal endpoints, which are pushed to infinitely remote times in this limit and do not constitute independent finite-frequency components of the photon distribution. They can therefore be omitted when expressing the asymptotic radiation spectrum in terms of the Fourier transform of the dipole.

The time dependence of the dipole operator can be expressed in terms of interaction-picture operators as
\begin{equation}
    \hat{\mathbf{d}}(t) =
    \hat{\mathbf{d}}_{\text{I}}(t)
    + i \int_{t_i}^{t} dt'' \,
    \left[
        \hat{\mathcal{V}}_{\text{I}}(t''),
        \hat{\mathbf{d}}_{\text{I}}(t)
    \right]
    +\cdots,
\end{equation}
where $\hat{\mathbf{d}}_{\text{I}}(t)$ and $\hat{\mathcal{V}}_{\text{I}}(t)$ are the dipole operator and light-matter interaction Hamiltonian, respectively, in the interaction picture defined with respect to the unperturbed Hamiltonian $\hat{H}_0$. The free evolution of the dipole is therefore given by
\begin{equation}
    \hat{\mathbf{d}}_{\mathrm{I}}(t)
    =
    \hat{U}_0^\dagger(t,t_i)\,
    \hat{\mathbf{d}}\,
    \hat{U}_0(t,t_i)
    =
    \hat{U}_\text{t}^\dagger(t,t_i)\,
    \hat{\mathbf{d}}\,
    \hat{U}_\text{t}(t,t_i)
    ,
\end{equation}
where $\hat{U}_0(t,t_i)$ and $\hat{U}_{\mathrm{t}}(t,t_i)$ are the evolution operators generated by the unperturbed Hamiltonian $\hat{H}_0=\hat{H}_{\mathrm{t}}+\hat{H}_{\mathrm{b}}$ and the target Hamiltonian $\hat{H}_{\mathrm{t}}$, respectively. Since $\hat{H}_{\mathrm{t}}$ and $\hat{H}_{\mathrm{b}}$ act on distinct Hilbert spaces and commute, the unperturbed evolution factorizes as $\hat{U}_0=\hat{U}_{\mathrm{t}}\hat{U}_{\mathrm{b}}$, and the beam evolution cancels according to $ \hat{U}_{\mathrm{b}}^\dagger(t,t_i) \hat{U}_{\mathrm{b}}(t,t_i) = \hat{1}$.

Using the electric-dipole form of the light-matter interaction, $\hat{\mathcal{V}}_{\mathrm{I}}(t) = -\hat{\mathbf{d}}_{\mathrm{I}}(t) \cdot \hat{\mathbf{E}}_{\mathrm{I}}(t)$, the perturbative expansion of the dipole operator takes the form
\begin{equation}
\label{eq:dip_op_t}
    \hat{\mathbf{d}}(t)
    =
    \hat{\mathbf{d}}_{\mathrm{I}}(t)
    +
    i\int_{t_i}^{t}dt''\,
    \left[
        \hat{\mathbf{d}}_{\mathrm{I}}(t),
        \hat{\mathbf{d}}_{\mathrm{I}}(t'') \cdot \hat{\mathbf{E}}_{\mathrm{I}}(t'')
    \right]
    +\cdots,
\end{equation}
where we have also exchanged the order of the operators in the commutator to obtain a form convenient for subsequent analysis. Substituting Eq.~(\ref{eq:dip_op_t}) into Eq.~(\ref{eq:ar_t_dip_approx}), we obtain
\begin{equation}
\label{eq:ar_dip_pert}
\begin{split}
    \hat{\mathrm a}^{\mathrm r}_{\mathbf k,\lambda}(t)
    = &
    \sqrt{\frac{2\pi\omega_{\mathbf k}}{V}}
    \int_{t_i}^{t}dt'\,
    e^{-i\omega_{\mathbf k}(t-t')}
    \boldsymbol{\epsilon}_{\mathbf k,\lambda}^{*}
    \cdot
    \hat{\mathbf d}_{\mathrm I}(t')
    \\
    & +
    i\sqrt{\frac{2\pi\omega_{\mathbf k}}{V}}
    \int_{t_i}^{t}dt'\,
    \int_{t_i}^{t}dt''\,
    e^{-i\omega_{\mathbf k}(t-t')}
    \Theta(t'-t'')\,
    \boldsymbol{\epsilon}_{\mathbf k,\lambda}^{*}
    \cdot
    \left[
        \hat{\mathbf d}_{\mathrm I}(t'),
        \hat{\mathbf d}_{\mathrm I}(t'')
        \cdot \hat{\mathbf E}_{\mathrm I}(t'')
    \right]
    +\cdots,
\end{split}
\end{equation}
where we introduced the Heaviside step function $\Theta(t'-t'')$ to extend the upper bound of the integration domain of $t''$ from $t'$ to $t$ while retaining the causal time ordering $t''\leq t'$. The first term in Eq.~(\ref{eq:ar_dip_pert}) describes the radiation generated by the freely evolving target dipole, whereas the second term describes the leading field-induced contribution.

The dipole operator and the electromagnetic field operator act on different subspaces and therefore commute. Consequently, the target radiation induced by the probe field, to linear order in the field, can be written as 
\begin{equation}
    \hat{\mathrm a}^{\mathrm r}_{\mathbf k,\lambda}(t) = 
    i\sqrt{\frac{2\pi\omega_{\mathbf k}}{V}}
    \int_{t_i}^{t}dt'\,
    \int_{t_i}^{t}dt''\,
    e^{-i\omega_{\mathbf k}(t-t')}
    \Theta(t'-t'')
    \boldsymbol{\epsilon}_{\mathbf k,\lambda}^{*}
    \cdot
    \hat{\boldsymbol{\mathcal D}}(t',t'')
    \cdot
    \hat{\mathbf E}_{\mathrm I}(t''),
\end{equation}
where $\hat{\boldsymbol{\mathcal D}}(t',t'')$ denotes the dipole-dipole commutator tensor, with elements
\begin{equation}
    \hat{\mathcal D}_{ij}(t',t'')
    =
    \left[
        \hat{d}^{i}_{\mathrm I}(t'),
        \hat{d}^{j}_{\mathrm I}(t'')
    \right],
\end{equation}
and $\hat{d}^{i}_{\mathrm I}(t)$ is the $i$-th component of the target dipole operator vector in the interaction picture at time $t$. Thus, the linear response of the target to the quantized probe field is determined by the dipole-dipole commutator of the field-free system, thereby permitting convenient separation of the target and the electromagnetic field.

\section{Evaluation of the electric susceptibility tensor}
\label{app:susceptibility}

We evaluate here the two-time electric susceptibility tensor
\begin{equation}
    \chi_{ij}(t',t'')
    =
    -i\Theta(t'-t'')
    \operatorname{Tr}
    \left\{
        \hat{\rho}_{\mathrm t}(t_i)
        \left[
            \hat d^{i}_{\mathrm I}(t'),
            \hat d^{j}_{\mathrm I}(t'')
        \right]
    \right\},
\end{equation}
where the trace is taken over the target degrees of freedom. Since the dipole and density operators are Hermitian, the commutator can be written as
\begin{equation}
    \chi_{ij}(t',t'')
    =
    2\Theta(t'-t'')
    \operatorname{Im}
    \operatorname{Tr}
    \left\{
        \hat{\rho}_{\mathrm t}(t_i)
        \hat d^{i}_{\mathrm I}(t')
        \hat d^{j}_{\mathrm I}(t'')
    \right\}.
\end{equation}

Assuming a pure initial target state, we expand the target density operator in the energy eigenbasis of $\hat H_{\mathrm t}$, $\hat H_{\mathrm t}|n\rangle = E_n|n\rangle$, as
\begin{equation}
    \hat{\rho}_{\mathrm t}(t_i)
    =
    \sum_{m,n}
    \rho_{nm}
    |n\rangle\langle m|,
\end{equation}
where $\rho_{nm}$ are, in general, complex matrix elements specifying the populations and coherences of the initial state. In the same basis, the interaction-picture dipole operator is
\begin{equation}
    \hat d^{i}_{\mathrm I}(t)
    =
    \sum_{l,k}
    d^{i}_{lk}
    e^{i(E_l-E_k)(t-t_i)}
    |l\rangle\langle k|,
\end{equation}
where $d^{i}_{lk}$ denotes the $i$-th component of the dipole matrix element, $\mathbf d_{lk}=\langle l|\hat{\mathbf d}|k\rangle$.

Substituting the above expressions and collecting the phase factors yields
\begin{equation}
    \chi_{ij}(t',t'')
    =
    2\Theta(t'-t'')
    \operatorname{Im}
    \left[
        \sum_{m,n,l}
        \rho_{nm}\,
        d^i_{ml}d^j_{ln}
        e^{-i(E_m-E_n)t_i}
        e^{i(E_m-E_l)t'}
        e^{i(E_l-E_n)t''}
    \right],
\end{equation}
which provides the spectral representation of the two-time susceptibility for an arbitrary coherent initial state of the target.

Let us evaluate the integral
\begin{equation}
    \mathcal{I}_{ij}(\omega,\omega_\mathbf{k}) = 
        \int_{-\infty}^{+\infty} dt' \int_{-\infty}^{+\infty} dt''
        \chi_{ij}(t',t'')
        e^{i\omega t'}e^{-i\omega_{\mathbf{k}}t''},
\end{equation}
which is the key quantity for obtaining a practical equation for computing the transient absorption spectra. Using $\operatorname{Im}X=-i(X-X^*)/2$ and introducing the finite lifetime of the intermediate states by replacing $E_l=E_l - i \Gamma_l/2$, we obtain
\begin{equation}
\begin{split}
    \mathcal{I}_{ij}(\omega,\omega_{\mathbf{k}})
    =& -i\sum_{m,n,l}
    \rho_{nm}d^i_{ml}d^j_{ln}
    e^{-i(E_m-E_n)t_i}
    \int_{-\infty}^{+\infty}dt'
    \int_{-\infty}^{t'}dt''
    e^{[i(E_m-E_l+\omega)-\Gamma_l/2]t'}
    e^{[i(E_l-E_n-\omega_{\mathbf{k}})+\Gamma_l/2]t''}
    \\
    &+i\sum_{m,n,l}
    \rho_{mn}d^j_{nl}d^i_{lm}
    e^{i(E_m-E_n)t_i}
    \int_{-\infty}^{+\infty}dt'
    \int_{-\infty}^{t'}dt''
    e^{[-i(E_m-E_l-\omega)-\Gamma_l/2]t'}
    e^{[-i(E_l-E_n+\omega_{\mathbf{k}})+\Gamma_l/2]t''}.
\end{split}
\end{equation}
Evaluating the inner integrals and using $\int_{-\infty}^{+\infty} dt e^{i\omega t}=2\pi \delta(\omega)$, we obtain
\begin{equation}
\begin{split}
    \mathcal{I}_{ij}(\omega,\omega_{\mathbf{k}})
    =& -2\pi\sum_{m,n,l}
    \frac{\rho_{nm}d^i_{ml}d^j_{ln}e^{-i(E_m-E_n)t_i}}
    {E_l-E_n-\omega_{\mathbf{k}}-i\Gamma_l/2}
    \delta\left(\omega-\omega_{\mathbf{k}}+E_m-E_n\right)
    \\
    &-2\pi\sum_{m,n,l}
    \frac{\rho_{mn}d^j_{nl}d^i_{lm}e^{i(E_m-E_n)t_i}}
    {E_l-E_n+\omega_{\mathbf{k}}+i\Gamma_l/2}
    \delta\left(\omega-\omega_{\mathbf{k}}-E_m+E_n\right).
\end{split}
\end{equation}
Relabeling $m\leftrightarrow n$ in the second term yields
\begin{equation}
\begin{split}
    \mathcal{I}_{ij}(\omega,\omega_{\mathbf{k}})
    = -2\pi \sum_{m,n,l}
    \rho_{nm}\,e^{-i(E_m-E_n)t_i}
    \delta\left(\omega-\omega_{\mathbf{k}}+E_m-E_n\right)
    \left[
    \frac{d^i_{ml}d^j_{ln}}
         {E_l-E_n-\omega_{\mathbf{k}}-i\Gamma_l/2}
    +
    \frac{d^j_{ml}d^i_{ln}}
         {E_l-E_m+\omega_{\mathbf{k}}+i\Gamma_l/2}
    \right].
\end{split}
\end{equation}
Thus, the double time integral is reduced to a sum over intermediate states with Lorentzian broadening $\Gamma_l$ and an energy-conserving delta function relating the frequencies $\omega$ and $\omega_{\mathbf{k}}$ and the energy difference $E_m-E_n$ between the corresponding states of the target.

\section{Multimode squeezed light}
\label{app:squeezing}

We consider here a general multimode squeezed optical state of the form
\begin{equation}
\label{eq:app_squeezed_state}
    |\Psi\rangle =
    \hat{S}(\boldsymbol{\beta})|0\rangle,
\end{equation}
where $|0\rangle$ denotes the electromagnetic vacuum state and $\hat{S}(\boldsymbol{\beta})$ is the multimode squeezing operator (see, e.g., Ref.~\cite{lo1993}),
\begin{equation}
\label{eq:app_squeezing_operator}
    \hat{S}(\boldsymbol{\beta})
    =
    \exp\left\{
        \frac{1}{2}
        \left(
            \hat{\mathbf{a}}^{\dagger}
                \boldsymbol{\beta}
            \hat{\mathbf{a}}^{\dagger T}
                -
            \hat{\mathbf{a}}^T
                \boldsymbol{\beta}^{*}
            \hat{\mathbf{a}}
        \right)
    \right\}.
\end{equation}
Here, $\hat{\mathbf{a}}=(...,\hat{a}_j,...)^{\text{T}}$ is the column vector of annihilation operators associated with the optical modes $j=(\mathbf{k},\lambda)$, $\hat{\mathbf{a}}^T$ is its row counterpart, $\hat{\mathbf{a}}^{\dagger}=(...,\hat{a}_j^{\dagger},...)$ is the row vector of creation operators and $\hat{\mathbf{a}}^{\dagger T}$ its column counterpart. The matrix $\boldsymbol{\beta}$ is complex symmetric, $\boldsymbol{\beta}^{\text{T}}=\boldsymbol{\beta}$, and specifies the magnitude and phase of the squeezing correlations between the optical modes. Both terms in the exponent are scalars and form an anti-Hermitian operator, so that $\hat{S}(\boldsymbol{\beta})$ is unitary.

Any complex symmetric matrix can be decomposed using the Autonne--Takagi factorization as
\begin{equation}
\label{eq:decomposition}
    \boldsymbol{\beta}
    =
    \mathbf{U}
    \mathbf{R}
    \mathbf{U}^{\mathrm{T}},
\end{equation}
where $\mathbf{U}$ is a complex unitary matrix and $\mathbf{R}$ is a real, non-negative diagonal matrix, defined as
\begin{equation}
    \mathbf{R} =
    \operatorname{diag}(R_1,R_2,\ldots),
    \qquad
    R_j \geq 0.
\end{equation}
Using the decomposition in Eq.~(\ref{eq:decomposition}), the multimode squeezing operator can be written as
\begin{equation}
\label{eq:S_factorized}
    \hat{S}(\boldsymbol{\beta})
    =
    \exp\left\{
        \frac{1}{2}
        \left(
            \hat{\mathbf{b}}^{\dagger}
            \mathbf{R}
            \hat{\mathbf{b}}^{\dagger T}
            -
            \hat{\mathbf{b}}^{\mathrm{T}}
            \mathbf{R}
            \hat{\mathbf{b}}
        \right)
    \right\} = 
    \exp\left\{
        \sum_j \hat{G}_j
    \right\} =
    \prod_j \exp\left\{
        \hat{G}_j
    \right\},
\end{equation}
where
\begin{equation}
    \hat{G}_j =
    \frac{R_j}{2} \left(
        [\hat{b}_j^{\dagger}]^2 - \hat{b}_j^2
    \right).
\end{equation}
Here, we have introduced the collective-mode annihilation and creation operators,
\begin{equation}
    \hat{\mathbf{b}} = \mathbf{U}^{\dagger} \hat{\mathbf{a}}
        \qquad \text{and} \qquad
    \hat{\mathbf{b}}^{\dagger} = \hat{\mathbf{a}}^{\dagger} \mathbf{U},
\end{equation}
in which the multimode squeezing transformation is diagonal. As for the original modes, $\hat{\mathbf{b}}$ is a column vector and $\hat{\mathbf{b}}^{\dagger}$ a row vector, so that $\hat{\mathbf{b}}^{\dagger T}$ denotes the column vector of the collective creation operators. Each collective mode is, in general, a coherent superposition of the original modes with different wave vectors and polarizations, while the corresponding diagonal element $R_j$ determines its squeezing strength.

Evaluating the expectation value of an operator $\hat{O}$ with respect to the state in Eq.~(\ref{eq:app_squeezed_state}), we can write
\begin{equation}
    \langle \hat{O}\rangle
    =
    \langle\Psi|\hat{O}|\Psi\rangle
    =
    \langle0|
    \hat{S}^{\dagger}\hat{O}\hat{S}
    |0\rangle.
\end{equation}
For $\hat{O}=\hat{b}_j$, since the multimode squeezing operator factorizes, Eq.~(\ref{eq:S_factorized}), and the generators $\hat{G}_k$ act on independent collective modes, we have
\begin{equation}
    \hat{S}^{\dagger}\hat{b}_j\hat{S} =
    e^{-\hat{G}_j}\hat{b}_j e^{\hat{G}_j} =
        \hat{b}_j
        +
        [\hat{b}_j,\hat{G}_j]
        +
        \frac{1}{2!}
        [[\hat{b}_j,\hat{G}_j],\hat{G}_j]
        +\cdots,
\end{equation}
where we utilized the Baker--Campbell--Hausdorff expansion. Using the bosonic commutation relation $[\hat{b}_j,\hat{b}_k^\dagger]=\delta_{jk}$, the required commutators can be written as
\begin{align}
    [\hat{b}_j,\hat{G}_j]
    & =
    R_j\hat{b}_j^\dagger, \\
    [[\hat{b}_j,\hat{G}_j],\hat{G}_j]
    &=
    R_j^2\hat{b}_j,\\
    [[[ \hat{b}_j,\hat{G}_j],\hat{G}_j],\hat{G}_j]
    &=
    R_j^3\hat{b}_j^\dagger.
\end{align}
Therefore, we obtain
\begin{equation}
\label{eq:SbS}
    \hat{S}^{\dagger}\hat{b}_j\hat{S} =
    \hat{b}_j
        \left(
            1+\frac{R_j^2}{2!}+\frac{R_j^4}{4!}+\cdots
        \right)
    +
    \hat{b}_j^\dagger
        \left(
            R_j+\frac{R_j^3}{3!}+\frac{R_j^5}{5!}+\cdots
        \right)
    = \hat{b}_j\cosh R_j + \hat{b}_j^\dagger\sinh R_j,
\end{equation}
and, proceeding in the same way for the creation operators,
\begin{equation}
\label{eq:Sb+S}
    \hat{S}^{\dagger}\hat{b}_j^\dagger\hat{S} =
    \hat{b}_j^\dagger\cosh R_j
    +
    \hat{b}_j\sinh R_j.
\end{equation}
Returning to the vector notation, the two relations read
\begin{equation}
\label{eq:SbS_matrix}
    \hat{S}^{\dagger}\hat{\mathbf{b}}\hat{S} =
    \cosh(\mathbf{R})\, \hat{\mathbf{b}}
    +
    \sinh(\mathbf{R})\, \hat{\mathbf{b}}^{\dagger T},
    \qquad
    \hat{S}^{\dagger}\hat{\mathbf{b}}^{\dagger T}\hat{S} =
    \cosh(\mathbf{R})\, \hat{\mathbf{b}}^{\dagger T}
    +
    \sinh(\mathbf{R})\, \hat{\mathbf{b}},
\end{equation}
where $\cosh(\mathbf{R})$ and $\sinh(\mathbf{R})$ are diagonal matrices with the elements $\cosh R_j$ and $\sinh R_j$, respectively.

Since the vacuum expectation values of annihilation and creation operators vanish, $\langle 0| \hat{b}_j |0\rangle = \langle 0| \hat{b}^{\dagger}_j |0\rangle = 0$, Eq.~(\ref{eq:SbS_matrix}) immediately gives
\begin{equation}
\label{eq:b_exp_value}
    \langle\hat{\mathbf{b}}\rangle
    =
    \langle\hat{\mathbf{b}}^{\dagger T}\rangle
    =
    \mathbf{0}.
\end{equation}
The second-order moments follow by applying the same transformation to the outer products of the operator vectors, whose $(j,l)$ elements are the corresponding products of the mode operators. This gives
\begin{align}
    \langle\hat{\mathbf{b}}^{\dagger T}\hat{\mathbf{b}}^{T}\rangle
    &=
    \sinh^2(\mathbf{R}),
    \label{eq:bdagger_b}\\
    \langle\hat{\mathbf{b}}\hat{\mathbf{b}}^{\dagger}\rangle
    &=
    \cosh^2(\mathbf{R}),
    \label{eq:b_bdagger}\\
    \langle\hat{\mathbf{b}}\hat{\mathbf{b}}^{T}\rangle
    =
    \langle\hat{\mathbf{b}}^{\dagger T}\hat{\mathbf{b}}^{\dagger}\rangle
    &=
    \frac{1}{2}\sinh(2\mathbf{R}).
    \label{eq:b_b}
\end{align}
All these matrices are diagonal, so that the Autonne--Takagi factorization diagonalizes not only the squeezing operator but also the second-order correlations of the squeezed vacuum.

Transforming these correlations back to the original optical modes, however, generally produces nonzero correlations between different wave vectors and polarizations. Inverting the definition of the collective operators, we obtain
\begin{equation}
\label{eq:a_from_b}
    \hat{\mathbf{a}} = \mathbf{U} \hat{\mathbf{b}}
        \qquad \text{and} \qquad
    \hat{\mathbf{a}}^{\dagger} = \hat{\mathbf{b}}^{\dagger} \mathbf{U}^{\dagger}.
\end{equation}
Since the first moments of the collective operators vanish, Eq.~(\ref{eq:b_exp_value}), the mean field amplitudes of the squeezed vacuum vanish as well,
\begin{equation}
\label{eq:a_first_moments}
    \langle\hat{\mathbf{a}}\rangle
    =
    \langle\hat{\mathbf{a}}^{\dagger T}\rangle
    =
    \mathbf{0},
\end{equation}
and, consequently, so does the expectation value of the electric field operator. Substituting Eq.~(\ref{eq:a_from_b}) into Eqs.~(\ref{eq:bdagger_b})--(\ref{eq:b_b}), the second-order moments in the original mode basis become
\begin{align}
    \langle\hat{\mathbf{a}}^{\dagger T}\hat{\mathbf{a}}^{T}\rangle
    &=
    \mathbf{U}^{*}\sinh^2(\mathbf{R})\,\mathbf{U}^{T},
    \label{eq:adagger_a}\\
    \langle\hat{\mathbf{a}}\hat{\mathbf{a}}^{\dagger}\rangle
    &=
    \mathbf{U}\cosh^2(\mathbf{R})\,\mathbf{U}^{\dagger},
    \label{eq:a_adagger}\\
    \langle\hat{\mathbf{a}}\hat{\mathbf{a}}^{T}\rangle
    &=
    \frac{1}{2}\mathbf{U}\sinh(2\mathbf{R})\,\mathbf{U}^{T},
    \label{eq:a_a}\\
    \langle\hat{\mathbf{a}}^{\dagger T}\hat{\mathbf{a}}^{\dagger}\rangle
    &=
    \frac{1}{2}\mathbf{U}^{*}\sinh(2\mathbf{R})\,\mathbf{U}^{\dagger}.
    \label{eq:adagger_adagger}
\end{align}
Equations~(\ref{eq:adagger_a}) and~(\ref{eq:adagger_adagger}) are the normal and the anomalous correlations entering the field-dependent spectral factor, Eq.~(\ref{eq:SF_full}), while the diagonal of Eq.~(\ref{eq:adagger_a}) gives the mean photon number in each optical mode.

For the two-mode squeezed vacuum used in Sec.~\ref{sec:sqp-picture}, the chosen squeezing matrix in the basis of the collective modes $A$ and $B$ is
\begin{equation}
    \boldsymbol{\beta}
    =
    R
    \begin{pmatrix}
        0 & 1\\
        1 & 0
    \end{pmatrix},
\end{equation}
for which the Autonne--Takagi factorization gives
\begin{equation}
    \mathbf{U}
    =
    \frac{1}{\sqrt{2}}
    \begin{pmatrix}
        1 & i\\
        1 & -i
    \end{pmatrix},
    \qquad
    \mathbf{R} = \operatorname{diag}(R,R),
\end{equation}
and therefore
\begin{equation}
    \langle\hat{\mathbf{a}}^{\dagger T}\hat{\mathbf{a}}^{T}\rangle
    =
    \sinh^2\!R
    \begin{pmatrix}
        1 & 0\\
        0 & 1
    \end{pmatrix},
    \qquad
    \langle\hat{\mathbf{a}}^{\dagger T}\hat{\mathbf{a}}^{\dagger}\rangle
    =
    \frac{1}{2}\sinh(2R)
    \begin{pmatrix}
        0 & 1\\
        1 & 0
    \end{pmatrix}.
\end{equation}
The two collective modes therefore carry equal mean photon numbers $\sinh^2 R$ and exhibit no normal correlations between them, whereas their anomalous correlations are $1/2 \sinh(2R)$, with the diagonal anomalous correlations vanishing.

\end{widetext}

\bibliographystyle{apsrev4-2}
\bibliography{full.bib}

\end{document}